\documentclass[10pt,journal,final,twocolumn]{IEEEtran}
\usepackage{graphicx}
\usepackage{color}
\usepackage{placeins}
\usepackage{float}
\usepackage{tabularx,colortbl}
\usepackage{amssymb}
\usepackage{amsthm}
\usepackage{cite}
\usepackage{amsmath}
\usepackage{bm}

\usepackage{caption2}

\usepackage{algorithm}
\usepackage{algpseudocode}
\theoremstyle{plain}
\newtheorem{thm}{Theorem}%%[section]

\theoremstyle{plain}
\newtheorem{rem}{Remark}
\newtheorem{cor}{Corollary}

\IEEEoverridecommandlockouts

\begin{document}

%----------------------------title&author&thanks----------------------------

%\title{Cell-Free Massive MIMO With Rate-Splitting Over Spatially Correlated Rician Channels}
%\title{Rate-Splitting for Cell-Free Massive MIMO With Spatially Correlated Rician Channels}
\title{Rate-Splitting for Cell-Free Massive MIMO: Performance Analysis and Generative AI Approach}
\author{Jiakang~Zheng, Jiayi~Zhang,~\IEEEmembership{Senior Member}, Hongyang~Du, Ruichen~Zhang, \\ Dusit~Niyato,~\IEEEmembership{Fellow,~IEEE}, Octavia A. Dobre,~\IEEEmembership{Fellow,~IEEE} and Bo~Ai,~\IEEEmembership{Fellow,~IEEE}

%\thanks{This work was supported in part by the Talent Fund of Beijing Jiaotong University, in part by National Natural Science Foundation of China under Grant 62471027 \& 62221001, in part by Natural Science Foundation of Jiangsu Province, Major Project under Grant BK20212002, in part by the Fundamental Research Funds for the Central Universities under Grant 2022JBQY004, and in part by ZTE Industry-University-Institute Cooperation Funds under Grant No. IA20240319002 \& No. IA20240709018. The work of Dusit Niyato was supported by the National Research Foundation, Singapore, and Infocomm Media Development Authority under its Future Communications Research \& Development Programme, Defence Science Organisation (DSO) National Laboratories under the AI Singapore Programme (FCP-NTU-RG-2022-010 and FCP-ASTAR-TG-2022-003), Singapore Ministry of Education (MOE) Tier 1 (RG87/22), and the NTU Centre for Computational Technologies in Finance (NTU-CCTF). The work of Octavia A. Dobre was supported in part by the Natural Sciences and Engineering Research Council of Canada (NSERC) through its Discovery program. (\emph{Corresponding author: Jiayi Zhang and Jiakang Zheng.})}

\thanks{J. Zheng, and J. Zhang, are with the School of Electronic and Information Engineering, Beijing Jiaotong University, Beijing 100044, China, and also with the Frontiers Science Center for Smart High-speed Railway System, Beijing Jiaotong University, Beijing 100044, China (e-mail: {jiakangzheng, jiayizhang}@bjtu.edu.cn).}
\thanks{H. Du is with the Department of Electrical and Electronic Engineering, University of Hong Kong, Pok Fu Lam, Hong Kong (e-mail: duhy@eee.hku.hk).}
\thanks{R. Zhang, and D. Niyato, are with the College of Computing and Data Science, Nanyang Technological University, Singapore (e-mail: {ruichen.zhang, dniyato}@ntu.edu.sg).}
\thanks{O. A. Dobre is with Faculty of Engineering and Applied Science, Memorial University, Canada (e-mail: odobre@mun.ca).}
\thanks{B. Ai is with the State Key Laboratory of Rail Traffic Control and Safety, Beijing Jiaotong University, Beijing 100044, China, and also with the Frontiers Science Center for Smart High-speed Railway System, and also with Henan Joint International Research Laboratory of Intelligent Networking and Data Analysis, Zhengzhou University, Zhengzhou 450001, China, and also with Research Center of Networks and Communications, Peng Cheng Laboratory, Shenzhen, China (e-mail: boai@bjtu.edu.cn).}

% \thanks{J. Zheng, J. Zhang, and B. Ai are with Beijing Jiaotong University; H. Du, R. Zhang, and D. Niyato are with Nanyang Technological University; O. A. Dobre is with Memorial University.}
%}
}

\maketitle
\vspace{-1cm}
%----------------------------abstract----------------------------
\begin{abstract}

Cell-free (CF) massive multiple-input multiple-output (MIMO) provides a ubiquitous coverage to user equipments (UEs) but it is also susceptible to interference. Rate-splitting (RS) effectively extracts data by decoding interference, yet its effectiveness is limited by the weakest UE. In this paper, we investigate an RS-based CF massive MIMO system, which combines strengths and mitigates weaknesses of both approaches. Considering imperfect channel state information (CSI) resulting from both pilot contamination and noise, we derive a closed-form expression for the sum spectral efficiency (SE) of the RS-based CF massive MIMO system under a spatially correlated Rician channel. Moreover, we propose low-complexity heuristic algorithms based on statistical CSI for power-splitting of common messages and power-control of private messages, and genetic algorithm is adopted as a solution for upper bound performance. Furthermore, we formulate a joint optimization problem, aiming to maximize the sum SE of the RS-based CF massive MIMO system by optimizing the power-splitting factor and power-control coefficient. Importantly, we improve a generative AI (GAI) algorithm to address this complex and non-convexity problem by using a diffusion model to obtain solutions. Simulation results demonstrate its effectiveness and practicality in mitigating interference, especially in dynamic environments.

\end{abstract}


\IEEEpeerreviewmaketitle

\section{Introduction}

The cell-free (CF) massive multiple-input multiple-output (MIMO) system is a huge leap forward over conventional cellular networks \cite{Ngo2017Cell} and has become a cutting-edge technology for next-generation mobile communications because of the capability to match the rapid growth rate of mobile traffic while also ensuring ubiquitous connectivity \cite{interdonato2019ubiquitous}. These benefits arise from the deployment architecture of the CF massive MIMO system that numerous geographically distributed access points (APs) are managed by a central processing unit (CPU) to provide massive antenna and macro-diversity gains \cite{zhang2020prospective}. Moreover, APs coherently serve user equipments (UEs) using spatial multiplexing on shared time-frequency resources, and this user-centric approach makes the UE unaware of the cell boundary \cite{zheng2023mobile}. Therefore, CF massive MIMO systems have drawn considerable interest from academic researchers. For instance, the authors in \cite{8379438} indicated that CF massive MIMO systems inherit both the channel hardening and favorable propagation characteristics of conventional cellular massive MIMO, thereby reducing the negative effects of signal fading and interference leakage.
Moreover, CF massive MIMO systems have been shown to achieve a 95\%-likely per-user SE five times higher than small-cell systems, owing to their superior interference management capability through joint signal processing among APs \cite{bjornson2019making}.
Furthermore, distributed resource allocation algorithms can be employed for transceiver design, power-control, and user scheduling in CF massive MIMO systems, significantly reducing complexity of centralized schemes \cite{9570126}.
Indeed, positive gains of coherent processing and distributed resource allocation in CF massive MIMO systems are heavily dependent on the high accuracy of channel state information (CSI), which is vulnerable to pilot contamination and noise \cite{9322468}. Therefore, it is crucial to explore solutions to the challenges due to imperfect CSI.

Recently, rate-splitting (RS) has been introduced to enhance system performance under imperfect CSI, especially in multi-user scenarios to manage interference \cite{7470942}.
Furthermore, RS integrates two existing technologies, i.e., treating interference as noise and decoding interference, demonstrating notable versatility and attracting many practical implementations \cite{mao2018rate,9831440}.
For example, an unmanned aerial vehicle (UAV) employs RS to facilitate massive access for Internet of Things (IoT) devices, achieving improved SE and reduced hardware complexity \cite{9324793}. Similarly, RS framework showed great potential in multibeam satellite communications for effectively suppressing inter-beam interference \cite{9257433}. Moreover, RS exhibits strong robustness to various non-ideal conditions, such as hardware impairment \cite{7892949}, Doppler issues \cite{9491092}, limited feedback \cite{7434643}, and asynchronous reception \cite{zheng2023mobile}. In the recent study by \cite{9942944}, it was shown that RS can mitigate pilot contamination and outperforms transmission without RS in terms of SE.
In addition, the common message of RS can be utilized as a jamming signal to enhance the covert communication performance \cite{10606453}.
Therefore, RS emerges as a key solution to challenges faced when applying CF massive MIMO in various complex scenarios. Meanwhile, uniform coverage capabilities of CF massive MIMO technologies enable RS to be used with a simple signal processing scheme to obtain high performance gains \cite{10032129}. An integration of CF massive MIMO and RS attracts a lot of interest. For instance, the authors in \cite{9737523} developed a effective max-min power-control algorithm and a heuristic common precoder for RS-based CF massive MIMO systems. Furthermore, a closed-form sum SE expression was derived for the RS-based CF massive MIMO system in \cite{10032129}, along with an introduction of a robust common precoder against asynchronous reception problems. Moreover, an achievable sum SE of RS-based CF massive MIMO systems was investigated in \cite{zheng2023rate}, highlighting their strong capability to mitigate channel aging effects.

However, most of the current research on RS-based CF massive MIMO systems considers Rayleigh channels \cite{9737523,10032129,zheng2023rate}, in which this assumption might not be practical for densely deployed networks. By deploying high-density APs, the distance between transceivers of CF massive MIMO reduces \cite{Ngo2017Cell}. As a result, channels are typically characterized as a superposition of a deterministic line-of-sight (LoS) path and spatially correlated Rayleigh fading caused by multipath propagation, which is known as spatially correlated Rician channels \cite{8809413}. Compared with non-LoS (NLoS) paths, which can experience increased fading due to scatterers, a LoS path typically provides a stronger and more stable signal, reducing channel estimation error \cite{8620255}. Hence, a proportion of a LoS component may have a strong impact on performance of RS-based CF massive MIMO systems. Another significant challenge is how RS-based CF massive MIMO systems perform adaptive resource allocation in dynamic environments, such as varying LoS components and moving APs and UEs. In fact, many works utilize convex optimization to solve power-control and precoding design problems in RS-based MIMO systems \cite{9324793,9257433,7892949,9491092,7434643}. 
However, with a large number of parameters, such as in RS-based CF massive MIMO systems, the problems become exceedingly challenging for conventional methods. Moreover, when optimization time taken is greater than the coherence time, a solution will be no longer valid.
A promising alternative is to utilize a model-free method, such as deep reinforcement learning (DRL), which has an ability to interact with dynamic environments and make real-time decisions \cite{10032267}.
For instance, the authors in \cite{9562976} proposed a DRL framework to quickly find optimal power allocation that maximizes sum-rate for RS-based MIMO networks, given dynamic and uncertainty of a communication channel.
Besides, multi-agent DRL was improved as a disruptive technology for addressing challenging combinatorial optimization problems in future wireless communications \cite{liu2024graph}.
However, DRL is sensitive to hyperparameters, often requiring adjustments based on specific tasks, and it may face challenges when capturing features from high-dimensional data.
On the other hand, generative AI (GAI) produces highly creative data and is applied to handle various tasks, including text, image, and audio \cite{jovanovic2022generative}.
In particular, the GAI based on the diffusion model can accurately capture the distribution of complex data because of its forward and reverse diffusion processes \cite{ho2020denoising}.
Therefore, it has superior generation and flexible implementation capabilities compared with other GAI methods, such as transformers, generative adversarial networks and variational autoencoders \cite{du2023beyond}.
Furthermore, the authors in \cite{10158526} used GAI with diffusion models to go beyond DRL in capturing complex patterns and relationships. This leads to high-performance, time-efficient, and fast-converging solutions in resource optimization problems\footnote{A large number of distributed APs in CF massive MIMO and the message-splitting characteristic of RS technology increase the number of system parameters. Therefore, the diffusion-based GAI method's ability to process high-dimensional data is highly suitable for resource optimization in this system\cite{yang2023diffusion}.}.

Motivated by the aforementioned observations, we investigate downlink performance of RS-based CF massive MIMO systems over spatially correlated Rician channels with imperfect CSI. In addition, low-complexity power-splitting factors and power-control coefficients are each designed based on statistical CSI to enhance data transmission of the system practically. Finally, we propose a GAI-based algorithm with diffusion models to solve the joint optimization problem of power-splitting factors and power-control coefficients to maximize the sum SE of the RS-based CF massive MIMO system. 
The major contributions of this work are listed below:
\begin{itemize}
  \item Considering an impact of imperfect CSI due to pilot contamination and noise, and utilizing the use-and-then-forget (UatF) capacity bound \cite{bjornson2017massive}, we derive a closed-form expression for the sum SE of the RS-based CF massive MIMO system under a spatially correlated Rician channel. Furthermore, the achievable sum SE is calculated to confirm that our derived result serves as a tight lower bound. From the expression, it is found that the LoS component amplifies both the desired signal and multi-user interference, thereby necessitating the use of RS to decode interference.
  \item To improve practicality of RS-based CF massive MIMO, we propose low-complexity heuristic algorithms for power-splitting in common messages and power-control in private messages based on statistical CSI. Furthermore, time-consuming genetic algorithms are compared as upper bounds \cite{katoch2021review}. It is found that the heuristic algorithm performs better without LoS paths than with LoS paths and is well-suited for efficiently obtaining solutions in large-scale networks.
  \item We then develop GAI for the RS-based CF massive MIMO system, in which optimal power-splitting factors and power-control coefficients are optimized by a diffusion model. The results show that, compared with the DRL-based algorithm \cite{10158526}, our proposed GAI-based algorithm can smoothly and continuously converge to a better performance. Particularly, in dynamic environments, characterized by varying LoS components and spatial correlation in a channel, the GAI-based algorithm exhibits consistent high performance.
\end{itemize}

\textbf{Notation:} Column vectors are denoted by boldface lowercase letters \(\mathbf{x}\), while matrices are represented by boldface uppercase letters \(\mathbf{X}\). The \(n \times n\) identity matrix is symbolized by \(\mathbf{I}_n\). Superscripts \(x^\mathrm{*}\), \(\mathbf{x}^\mathrm{T}\), and \(\mathbf{x}^\mathrm{H}\) indicate the conjugate, transpose, and conjugate transpose operations, respectively. The notation \(\mathrm{diag}(x_1, \ldots, x_n)\) is used for a block-diagonal matrix where the diagonal entries are \(x_1, \ldots, x_n\). The symbols \(\left| \cdot \right|\), \(\left\| \cdot \right\|\), \(\mathbb{E}\{\cdot\}\), \(\text{tr}(\cdot)\), and \(\triangleq\) are used to denote the absolute value, Euclidean norm, expectation operator, trace operator, and definition, respectively. Lastly, \(\mathbf{x} \sim \mathcal{C}\mathcal{N}(\mathbf{0}, \mathbf{R})\) describes a circularly symmetric complex Gaussian distribution with covariance matrix \(\mathbf{R}\).

%----------------------------system model----------------------------
\section{System Model}\label{se:model}

Considering a CF massive MIMO system comprising $K$ UEs and $L$ APs, where each UE employs a single antenna while each AP uses $N$ antennas. All the $K$ UEs are simultaneously served by all the $L$ APs using the same time-frequency resources. Furthermore, we adopt a time-division duplex protocol using a standard coherence block model, which includes $\tau_c$ channel uses\footnote{The number of practically useful channel uses per coherence block can be less than the product of coherence time and coherence bandwidth \cite{bjornson2017massive}.}.
Moreover, $\tau_p$ channel uses are used as an uplink training phase and $\tau_c - \tau_p$ channel uses are used as a downlink transmission phase \cite{bjornson2019making}. Then, the channel ${{\mathbf{g}}_{kl}} \in {\mathbb{C}^{N \times 1}}$ between UE $k$ and AP $l$ is modeled as Rician fading:
\begin{align}
{{\mathbf{g}}_{kl}} \sim \mathcal{C}\mathcal{N}\left( {{{{\mathbf{\bar h}}}_{kl}},{{\mathbf{R}}_{kl}}} \right),
\end{align}
where the mean ${{{\mathbf{\bar h}}}_{kl}}\in {\mathbb{C}^{N\times 1}}$ is a vector denoting the LoS component, and the variance ${{\mathbf{R}}_{kl}} \in {\mathbb{C}^{N \times N}}$ is a matrix that depicts the spatial correlation between NLoS components. Furthermore, assuming that each AP uses a uniform linear array of omni-directional antennas, the LoS and NLoS components of the channel are modeled as follows \cite{8620255}:
\begin{align}
{{\mathbf{\bar h}}_{kl}} =& \sqrt {\beta _{kl}^{{\text{los}}}} {\left[ {1,{e^{j2\pi {d_{\text{H}}}\sin \left( {{\varphi _{kl}}} \right)}}, \ldots ,{e^{j2\pi {d_{\text{H}}}(N - 1)\sin \left( {{\varphi _{kl}}} \right)}}} \right]^{\text{T}}} , \notag \\
{\left[ {{{\mathbf{R}}_{kl}}} \right]_{s,m}} &\!=\! \frac{{\beta _{kl}^{{\text{nlos}}}}}{K}\!\sum\limits_{t = 1}^{{N_{\text{c}}}} \! {{e^{j\pi (s - m)\sin \left( {{\varphi _{kl,t}}} \right) - \frac{{\sigma _\varphi ^2}}{2}{{\left( {\pi (s - m)\cos \left( {{\varphi _{kl,t}}} \right)} \right)}^2}}}} ,  \notag
\end{align}
where ${{\varphi _{kl}}}$ is the angle of arrival (AoA) at which the LoS signal is transmitted from AP $l$ to UE $k$. Besides, ${\left[  \cdot  \right]_{s,m}}$ denotes the $\left( {s,m} \right)$th element of the matrix, and $N_\text{c}$ is the number of scattering clusters of NLoS signal. The term ${\varphi _{kl,t}}$ is the nominal AoA for cluster $t$, following a uniform distribution $\mathcal{U}\left[ {{\varphi _{kl}} - {40^\text{o} },{\varphi _{kl}} + {40^\text{o} }} \right]$ \cite{8620255}. Moreover, the AoAs of the multipath components within a cluster are Gaussian distributed around the nominal AoA, with an angular standard deviation (ASD) of ${\sigma _\varphi ^2}$ \cite{bjornson2017massive}.
Moreover, ${{d_{\text{H}}}} \leqslant 0.5$ denotes that the antenna spacing is half a wavelength or less. Furthermore, ${\beta _{kl}^{{\text{los}}}}$ and ${\beta _{kl}^{{\text{nlos}}}}$ refer to the large-scale fading coefficients for LoS and NLoS transmission paths from AP $l$ to UE $k$, and their values are calculated as
\begin{align}
  \beta _{kl}^{{\text{los}}} &= \frac{1}{N}{\left\| {{{{\mathbf{\bar h}}}_{kl}}} \right\|^2} = \sqrt {{\bar{K}_{kl}}/\left( {{\bar{K}_{kl}} + 1} \right)} {\zeta _{kl}} , \\
  \beta _{kl}^{{\text{nlos}}} &= \frac{1}{N}{\text{tr}}\left( {{{\mathbf{R}}_{kl}}} \right) = \sqrt {1/\left( {{\bar{K}_{kl}} + 1} \right)} {\zeta _{kl}} ,
\end{align}
where ${\bar{K}_{kl}}$ represents the Rician factor, and $\zeta _{kl}$ represents the large-scale fading coefficient for the channel.

\subsection{Channel Estimation}

Channel estimation is performed through the utilization of uplink pilot transmissions. Specifically, $\tau_p$ orthogonal pilot signals ${\phi _t} \in {\mathbb{C}^{{\tau _p} \times 1}},t = 1, \ldots ,{\tau _p}$ are randomly assigned to UEs.
The pilot sequence is designed with unit-magnitude elements to maintain a consistent power level, which leads to ${\left\| {{\bm{\phi} _t}} \right\|^2} = {\tau _p}$ \cite{bjornson2017massive}.
Considering a network with more UEs than the number of orthogonal pilot signals, it is possible for different UEs to be assigned the pilot signal.
Let $t_k$ denote the pilot index for UE $k$, with ${\mathcal{P}_k} = \left\{ {i:{t_i} = {t_k}} \right\} \subset \left\{ {1, \ldots ,K} \right\}$ defined as the set of UEs that share pilot $t_k$.
Then, the pilot signal ${{\mathbf{Z}}_l} \in {\mathbb{C}^{N \times {\tau _p}}}$ received at AP $l$ is obtained as follows:
\begin{align}
{{\mathbf{Z}}_l} = \sum\limits_{i = 1}^K {\sqrt {{p_i}} } {{\mathbf{g}}_{il}}\bm{\phi} _{{t_i}}^{\text{T}} + {{\mathbf{N}}_l} ,
\end{align}
where $p_i \geqslant 0$ denotes the transmit power of UE $i$, while ${{\mathbf{N}}_l} \in {\mathbb{C}^{N \times {\tau _p}}}$ represents the noise matrix with entries following independent distributions as $\mathcal{C}\mathcal{N}\left( {{{0}},\sigma^2} \right)$.
For estimating ${{\mathbf{g}}_{kl}}$, we correlate the received signal at the AP with the corresponding normalized pilot\footnote{This step suppresses the interference from other received pilot signals in CF massive MIMO systems, achieving the same effect as combining in uplink data transmission \cite{9586055}.}, which produces the following:
\begin{align}
  {{\mathbf{z}}_{kl}} &= \frac{1}{{\sqrt {{\tau _p}} }}{{\mathbf{Z}}_l}\bm{\phi} _{{t_k}}^*
   = \frac{1}{{\sqrt {{\tau _p}} }}\sum\limits_{i \in {\mathcal{P}_k}} {\sqrt {{p_i}} {{\mathbf{g}}_{il}}\bm{\phi} _{{t_i}}^{\text{T}}\bm{\phi} _{{t_k}}^*}  + \frac{1}{{\sqrt {{\tau _p}} }}{{\mathbf{N}}_l}\bm{\phi} _{{t_k}}^* \notag \\
   &= \! \sum\limits_{i \in {\mathcal{P}_k}} \!\! {\sqrt {{p_i}{\tau _p}} {{\mathbf{g}}_{il}}}  \!+\! {{\mathbf{n}}_{{t_k}l}}  \!=\! \sum\limits_{i \in {\mathcal{P}_k}} \!\! {\sqrt {{p_i}{\tau _p}} \left( {{{{\mathbf{\bar h}}}_{il}} \!+\! {{\mathbf{h}}_{il}}} \right)}  \!+\! {{\mathbf{n}}_{{t_k}l}} ,
\end{align}
where ${{\mathbf{n}}_{{t_k}l}} \!\!\in\!\! {\mathbb{C}^N}$ is the resulting noise with distribution $\mathcal{C}\mathcal{N}\left( {{\mathbf{0}},{\sigma ^2}{{\mathbf{I}}_N}} \right)$.
Utilizing estimation theory, the minimum mean-squared error (MMSE) estimate of ${{\mathbf{g}}_{kl}}$ can be derived as follows \cite{8620255}:
\begin{align}\label{Eq:ghat}
  {{{\mathbf{\hat g}}}_{kl}} = {{{\mathbf{\bar h}}}_{kl}} + {{{\mathbf{\hat h}}}_{kl}}
   = {{{\mathbf{\bar h}}}_{kl}} + \sqrt {{p_k}{\tau _p}} {{\mathbf{R}}_{kl}}{{\mathbf{\Psi }}_{kl}}\left( {{{\mathbf{z}}_{kl}} - {{{\mathbf{\bar z}}}_{kl}}} \right) ,
\end{align}
where ${{{\mathbf{\bar z}}}_{kl}} \in {\mathbb{C}^{N \times 1}}$ and ${{\mathbf{\Psi }}_{kl}} \in {\mathbb{C}^{N \times N}}$ are 
\begin{align}
{{{\mathbf{\bar z}}}_{kl}} &= \sum\limits_{i \in {\mathcal{P}_k}} {\sqrt {{p_i}{\tau _p}} {{{\mathbf{\bar h}}}_{il}}}  , \\
{{\mathbf{\Psi }}_{kl}} &= {\left( {\sum\limits_{i \in {\mathcal{P}_k}} {{p_i}{\tau _p}{{\mathbf{R}}_{il}}}  + {\sigma ^2}{{\mathbf{I}}_N}} \right)^{ - 1}}  .
\end{align}
Furthermore, the distributions of the estimate and the estimate error are derived as
\begin{align}
\label{Eq:ghat_kl} {{{\mathbf{\hat g}}}_{kl}} &\sim \mathcal{C}\mathcal{N}\left( {{{{\mathbf{\bar h}}}_{kl}},{{\mathbf{Q}}_{kl}}} \right) , \\
\label{Eq:gerror_kl} {{{\mathbf{\tilde g}}}_{kl}} = {{\mathbf{g}}_{kl}} - {{{\mathbf{\hat g}}}_{kl}} &\sim \mathcal{C}\mathcal{N}\left( {{\mathbf{0}},{{\mathbf{C}}_{kl}} } \right) ,
\end{align}
with ${{\mathbf{Q}}_{kl}} \in {\mathbb{C}^{N \times N}}$ and ${{\mathbf{C}}_{kl}} \in {\mathbb{C}^{N \times N}}$ are 
\begin{align}
{{\mathbf{Q}}_{kl}} &= {p_k}{\tau _p}{{\mathbf{R}}_{kl}}{{\mathbf{\Psi }}_{kl}}{{\mathbf{R}}_{kl}} , \\
{{\mathbf{C}}_{kl}} &= {{\mathbf{R}}_{kl}} - {{\mathbf{Q}}_{kl}}  .
\end{align}
For notational convenience, we define ${{{\mathbf{\bar Q}}}_{kil}} \in {\mathbb{C}^{N \times N}}$ as
\begin{align}
{{{\mathbf{\bar Q}}}_{kil}} = {p_k}{\tau _p}{{\mathbf{R}}_{il}}{{\mathbf{\Psi }}_{kl}}{{\mathbf{R}}_{kl}}  .
\end{align}

\newcounter{mytempeqncnt}
\begin{figure*}[t!]
\normalsize
\setcounter{mytempeqncnt}{1}
\setcounter{equation}{18}
\begin{align}
\label{eq:SINRc} {\text{SINR}}_k^{\text{c}} &= \frac{{{p_{\text{d}}}{{\left| {\sum\limits_{l = 1}^L {\sqrt {{\rho _l}} {\mathbf{\hat g}}_{kl}^{\text{H}}{\mathbf{v}}_{{\text{c}},l}^{{\text{norm}}}} } \right|}^2}}}{{{p_{\text{d}}}\sum\limits_{l = 1}^L {{\rho _l}{{\left( {{\mathbf{v}}_{{\text{c}},l}^{{\text{norm}}}} \right)}^{\text{H}}}{{\mathbf{C}}_{kl}}{\mathbf{v}}_{{\text{c}},l}^{{\text{norm}}}}  \!+\! \frac{{{p_{\text{d}}}}}{K}\sum\limits_{i = 1}^K {{{\left| {\sum\limits_{l = 1}^L {\sqrt {\left( {1 \!-\! {\rho _l}} \right){\eta _{il}}} } {\mathbf{\hat g}}_{kl}^{\text{H}}{\mathbf{v}}_{il}^{{\text{norm}}}} \right|}^2}}  \!+\! \frac{{{p_{\text{d}}}}}{K}\sum\limits_{i = 1}^K {\sum\limits_{l = 1}^L {\left( {1 \!-\! {\rho _l}} \right){\eta _{il}}{{\left( {{\mathbf{v}}_{il}^{{\text{norm}}}} \right)}^{\text{H}}}{{\mathbf{C}}_{kl}}{\mathbf{v}}_{il}^{{\text{norm}}}} } \!+\! {\sigma ^2}}} , \\
\label{eq:SINRp} {\text{SINR}}_k^{\text{p}} &= \frac{{\frac{{{p_{\text{d}}}}}{K}{{\left| {\sum\limits_{l = 1}^L {\sqrt {\left( {1 - {\rho _l}} \right){\eta _{kl}}} } {\mathbf{\hat g}}_{kl}^{\text{H}}{\mathbf{v}}_{kl}^{{\text{norm}}}} \right|}^2}}}{{\frac{{{p_{\text{d}}}}}{K}\sum\limits_{i \ne k}^K {{{\left| {\sum\limits_{l = 1}^L {\sqrt {\left( {1 - {\rho _l}} \right){\eta _{il}}} } {\mathbf{\hat g}}_{kl}^{\text{H}}{\mathbf{v}}_{il}^{{\text{norm}}}} \right|}^2} + \frac{{{p_{\text{d}}}}}{K}\sum\limits_{i = 1}^K {\sum\limits_{l = 1}^L {\left( {1 - {\rho _l}} \right){\eta _{il}}{{\left( {{\mathbf{v}}_{il}^{{\text{norm}}}} \right)}^{\text{H}}}{{\mathbf{C}}_{kl}}{\mathbf{v}}_{il}^{{\text{norm}}}} }  + {\sigma ^2}} }} .
\end{align}
\setcounter{equation}{13}
\hrulefill
\vspace{-0.2cm}
\end{figure*}

\subsection{Rate-Splitting Strategy}

Due to pilot contamination and noise in channel estimation, our considered CF massive MIMO system has imperfect CSI. To address this, we suggest using RS for downlink data transmission.
As illustrated in Fig.~\ref{fig:CF_RS_GAI}, at each AP of RS-based CF massive MIMO systems, we split a message of each UE into a common sub-message and a private sub-message, i.e., ${W_k} = {W_{{\text{c}},k}} + {W_{{\text{p}},k}},\forall k$. These common sub-messages are then combined into a super common message $s_\text{c}$, while each private sub-message remains as an individual private message ${s_1}, \ldots ,{s_K}$. Finally, superposition coding is employed for transmitting all these messages simultaneously.
At the receiver side, each UE decodes the common message, with all the private messages regarded as noise. Following this, the decoded common message is removed from the received signal using SIC, given an assumption of error-free\footnote{Each UE performs SIC only once, enabling the assumption of error-free decoding for preliminary analytical results. However, imperfect SIC will be addressed in the future work through modeling the residual coefficient of common messages\cite{10517628}.} decoding \cite{mao2018rate}. Each UE then decodes its private message by treating other private messages as noise.
Utilizing the normalized precoding vector ${\mathbf{v}}_{{\text{c}},l}^{{\text{norm}}} \in {\mathbb{C}^{N \times 1}}$ with the constraint ${\left\| {{\mathbf{v}}_{{\text{c}},l}^{{\text{norm}}}} \right\|^2} \leqslant 1$ for common messages and the normalized precoding vector ${{\mathbf{v}}_{il}^{{\text{norm}}}} \in {\mathbb{C}^{N \times 1}}$ with the constraint ${\left\| {{\mathbf{v}}_{il}^{{\text{norm}}}} \right\|^2} \leqslant 1$ for private messages in the CF massive MIMO system with RS, we express the transmitted signal ${{\mathbf{x}}_l} \in {\mathbb{C}^{N \times 1}}$ from AP $l$ as
\begin{align}\label{transmited_signal}
{{\mathbf{x}}_l} = \sqrt {{p_{\text{d}}}{\rho _l}} {\mathbf{v}}_{{\text{c}},l}^{{\text{norm}}}{s_{\text{c}}} + \sqrt {\frac{{{p_{\text{d}}}\left( {1 - {\rho _l}} \right)}}{K}} \sum\limits_{i = 1}^K {{\mathbf{v}}_{il}^{{\text{norm}}}\sqrt {{\eta _{il}}} {s_i}} ,
\end{align}
where ${s_{\text{c}}} \sim \mathcal{C}\mathcal{N}\left( {0,1} \right)$ and ${s_i} \sim \mathcal{C}\mathcal{N}\left( {0,1} \right)$ are the common and private messages, respectively. Furthermore, $p_\text{d}$ is the maximum downlink transmit power of each AP. In addition, $0 \leqslant {\rho _l} \leqslant 1$ is the fraction of power allocated for transmitting common messages at each AP, and $0 \leqslant {\eta _{il}} \leqslant 1$ is power-control coefficients, which are chosen to satisfy the downlink power constraint as $\mathbb{E}\left\{ {{{\left\| {{{\mathbf{x}}_l}} \right\|}^2}} \right\} \leqslant {p_{\text{d}}}$.
Note that the optimal values for power-splitting factors and power-control coefficients are influenced by system parameters, including, but not limited to, the signal-to-noise ratio (SNR), the number of APs, and the number of UEs \cite{7470942}. Consequently, it is significant to fine-tune these factors and coefficients to fully exploit the capabilities of RS and CF massive MIMO for performance improvement.
Furthermore, we have
\begin{align}
\label{eq:vc_norm} {\mathbf{v}}_{{\text{c}},l}^{{\text{norm}}} &= \sqrt {{\mu _{{\text{c}},l}}} {{\mathbf{v}}_{{\text{c}},l}} = {{{{\mathbf{v}}_{{\text{c}},l}}}}/{{\sqrt {\mathbb{E}\left\{ {{{\left\| {{{\mathbf{v}}_{{\text{c}},l}}} \right\|}^2}} \right\}} }}, \\
\label{eq:vp_norm} {\mathbf{v}}_{il}^{{\text{norm}}} &= \sqrt {{\mu _{il}}} {{\mathbf{v}}_{il}} = {{{{\mathbf{v}}_{il}}}}/{{\sqrt {\mathbb{E}\left\{ {{{\left\| {{{\mathbf{v}}_{il}}} \right\|}^2}} \right\}} }},
\end{align}
where ${{\mu _{{\text{c}},l}}}$ and ${{\mu _{il}}}$ are the normalized coefficients for common precoding and private precoding, respectively.

\section{Downlink Data Transmission}\label{se:performance}

This section focuses on the performance of RS-based CF massive MIMO systems under spatially correlated Rician fading. The superposition-based common precoding and the MR private precoding are adopted to derive the novel closed-form sum SE expressions.
Furthermore, we separately investigate the design of power-splitting factor for common messages and the design of power-control coefficients for private messages.

\subsection{Sum Spectral Efficiency}

In the RS-based CF massive MIMO system with coherent transmission, the transmitted signals are divided into common and private messages, as in \eqref{transmited_signal}. Then, the signal received at UE $k$ is expressed as
\begin{align}\label{eq:r_k}
  {r_k} &= \sum\limits_{l = 1}^L {{\mathbf{g}}_{kl}^{\text{H}}{{\mathbf{x}}_l}}  + {w_k} = \sqrt {{p_{\text{d}}}} \sum\limits_{l = 1}^L {\sqrt {{\rho _l}} {\mathbf{g}}_{kl}^{\text{H}}{\mathbf{v}}_{{\text{c}},l}^{{\text{norm}}}{s_{\text{c}}}}  \notag \\
   &+ \sqrt {\frac{{{p_{\text{d}}}}}{K}} \sum\limits_{l = 1}^L {\sqrt {\left( {1 - {\rho _l}} \right){\eta _{kl}}} } {\mathbf{g}}_{kl}^{\text{H}}{\mathbf{v}}_{kl}^{{\text{norm}}}{s_k} \notag \\
   &+ \sqrt {\frac{{{p_{\text{d}}}}}{K}} \sum\limits_{i \ne k}^K {\sum\limits_{l = 1}^L {\sqrt {\left( {1 - {\rho _l}} \right){\eta _{il}}} } {\mathbf{g}}_{kl}^{\text{H}}{\mathbf{v}}_{il}^{{\text{norm}}}{s_i}}  + {w_k},
\end{align}
where $w_k \sim \mathcal{C}\mathcal{N}\left( {0,{\sigma ^2}} \right)$ is the receiver noise.
Assuming the channel estimate ${{{\mathbf{\hat g}}}_{kl}}$ is available to the UEs, we have the achievable sum SE as
\begin{align}\label{achievable_sum_SE}
  {\mathrm{Sum\;}}{{\mathrm{SE}}} &= \frac{\tau_c - \tau_p}{\tau_c}\mathbb{E}\left\{ {{{\min }_k}\left\{ {{\mathrm{SE}}_{k}^{\mathrm{c}}} \right\}} \right\} + \frac{\tau_c - \tau_p}{\tau_c}\sum\limits_{k = 1}^K {\mathbb{E}\left\{ {{\mathrm{SE}}_{k}^{\mathrm{p}}} \right\}}  \notag \\
   &= \frac{\tau_c - \tau_p}{\tau_c}\mathbb{E}\left\{ {{{\log }_2}\left( {1 + {{\min }_k}\left\{ {{\mathrm{SINR}}_{k}^{\mathrm{c}}} \right\}} \right)} \right\} \notag \\
   &+ \frac{\tau_c - \tau_p}{\tau_c}\sum\limits_{k = 1}^K {\mathbb{E}\left\{ {{{\log }_2}\left( {1 + {\mathrm{SINR}}_{k}^{\mathrm{p}}} \right)} \right\}} ,
\end{align}
where ${{\text{SINR}}_k^{\text{c}}}$ and ${{\text{SINR}}_k^{\text{p}}}$ are given in \eqref{eq:SINRc} and \eqref{eq:SINRp} at the top of this page, respectively.
\begin{rem}
Note that the sum SE expression in \eqref{achievable_sum_SE} holds for any normalized precoding vectors ${{\mathbf{v}}_{{\mathrm{c}},l}^{{\mathrm{norm}}}}$ and ${{\mathbf{v}}_{kl}^{{\mathrm{norm}}}}$, and it is a multi-antenna generalization of \cite{8809413} under downlink RS-based CF massive MIMO and an extension of \cite{10032129,zheng2023rate} to spatially correlated Rician channels. Possible choices for private messages are to use zero-forcing precoding \cite{9529197}, local MMSE precoding and centralized MMSE precoding \cite{zheng2023rate}, which have high computational complexity due to inverse operations of high-dimensional matrices. Moreover, the precoding vector for common messages can be chosen from random precoding \cite{9491092}, superposition-based precoding \cite{zheng2023rate}, or optimization-based precoding \cite{9831440}, where the last one has the best performance but the highest complexity.
\end{rem}
For further analysis, we utilize the UatF bound\footnote{In the UatF bound, only the part of desired signal captured via an average precoded channel, denoted by $\mathbb{E}\left\{ {{\text{DS}}} \right\}$, is treated as the true desired signal. The beamforming gain uncertainty of signal can thus be treated as an uncorrelated noise in the detection \cite{bjornson2017massive}.} to derive a closed-form expression for sum SE of the considered RS-based CF massive MIMO system. From \eqref{eq:vc_norm} and \eqref{eq:vp_norm}, we can written \eqref{eq:r_k} as
\setcounter{equation}{20}
\begin{align}\label{eq:r_c}
  r_k^{\text{c}} &= \sqrt {{p_{\text{d}}}} \underbrace {\sum\limits_{l = 1}^L {\sqrt {{\rho _l}{\mu _{{\text{c}},l}}} \mathbb{E}\left\{ {{\mathbf{g}}_{kl}^{\text{H}}{{\mathbf{v}}_{{\text{c}},l}}} \right\}{s_{\text{c}}}} }_{{\text{DS}}_k^{\text{c}}} \notag \\
   &+ \sqrt {{p_{\text{d}}}} \underbrace {\sum\limits_{l = 1}^L {\sqrt {{\rho _l}{\mu _{{\text{c}},l}}} {\mathbf{g}}_{kl}^{\text{H}}{{\mathbf{v}}_{{\text{c}},l}}{s_{\text{c}}}} }_{{\text{REC}}_k^{\text{c}}} - \sqrt {{p_{\text{d}}}} {\text{DS}}_k^{\text{c}} \notag \\
   &+ \sqrt {\frac{{{p_{\text{d}}}}}{K}} \sum\limits_{i = 1}^K {\underbrace {\sum\limits_{l = 1}^L {\sqrt {\left( {1 - {\rho _l}} \right){\eta _{il}}{\mu _{il}}} } {\mathbf{g}}_{kl}^{\text{H}}{{\mathbf{v}}_{il}}{s_i}}_{{\text{REC}}_{ki}^{\text{p}}}}  + {w_k} \hfill ,
\end{align}
where ${{\text{DS}}_k^{\text{c}}}$ denotes the desired common messages and ${{\text{REC}}_k^{\text{c}}}$ denotes the received common messages. Thus, the beamforming gain uncertainty of common messages can be expressed as ${{\text{REC}}_k^{\text{c}}} - {{\text{DS}}_k^{\text{c}}}$ \cite{Ngo2017Cell,bjornson2017massive}. Moreover, ${{\text{REC}}_{ki}^{\text{p}}}$ denotes the received private messages of UE $i$ at UE $k$.
After cancellation of the common messages, the received private messages of UE $k$ can be expressed as
\begin{align}\label{eq:r_p}
  r_k^{\text{p}} &= \sqrt {\frac{{{p_{\text{d}}}}}{K}} \underbrace {\sum\limits_{l = 1}^L {\sqrt {\left( {1 - {\rho _l}} \right){\eta _{kl}}{\mu _{kl}}} } \mathbb{E}\left\{ {{\mathbf{g}}_{kl}^{\text{H}}{{\mathbf{v}}_{kl}}} \right\}{s_k}}_{{\text{DS}}_k^{\text{p}}} \notag \\
   &+ \sqrt {\frac{{{p_{\text{d}}}}}{K}} \sum\limits_{i = 1}^K {{\text{REC}}_{ki}^{\text{p}}}  - \sqrt {\frac{{{p_{\text{d}}}}}{K}} {\text{DS}}_k^{\text{p}} + {w_k} ,
\end{align}
where ${{\text{DS}}_k^{\text{p}}}$ denotes the desired private messages of UE $k$, and then the beamforming gain uncertainty of private messages of UE $k$ can be expressed as ${\text{REC}}_{kk}^{\text{p}} - {\text{DS}}_k^{\text{p}}$ \cite{Ngo2017Cell,bjornson2017massive}.

\begin{thm}\label{thm:1}
Using MR precoding ${{\mathbf{v}}_{il}} = {{{\mathbf{\hat g}}}_{il}}$ for private messages and superposition-based precoding ${{\mathbf{v}}_{{\mathrm{c}},l}} = \sum\nolimits_{i = 1}^K {{{{\mathbf{\hat g}}}_{il}}}$ for common messages, a lower bound on the sum SE is
\begin{align}\label{eq:sumSE}
{\mathrm{Sum}}\;{\mathrm{SE}} &= \frac{\tau_c - \tau_p}{\tau_c}{\log _2}\left( {1 + {{\min }_k}\left\{ {{\mathrm{SINR}}_k^{\mathrm{c}}} \right\}} \right) \notag\\
&+ \frac{\tau_c - \tau_p}{\tau_c}\sum\limits_{k = 1}^K {{{\log }_2}\left( {1 + {\mathrm{SINR}}_k^{\mathrm{p}}} \right)},
\end{align}
with ${{\mathrm{SINR}}_k^{\mathrm{c}}}$ and ${\mathrm{SINR}}_k^{\mathrm{p}}$ are given by
\begin{align}
  {\mathrm{SINR}}_k^{\mathrm{c}} &\!=\! \frac{{{p_{\mathrm{d}}}{\mathrm{T}}_k^{{\mathrm{c}}1}}}{{{p_{\mathrm{d}}}{\mathrm{T}}_k^{{\mathrm{c}}2} \!+\! \frac{{{p_{\mathrm{d}}}}}{K}\!\sum\limits_{i = 1}^K \! {{\mathrm{T}}_{ki}^{{\mathrm{p}}2}}  \!+\! \frac{{{p_{\mathrm{d}}}}}{K}\!\sum\limits_{i \in {\mathcal{P}_k}}^K \!\! {{\mathrm{T}}_{ki}^{{\mathrm{p}}1}}  \!+\! \frac{{{p_{\mathrm{d}}}}}{K}\!\sum\limits_{i \notin {\mathcal{P}_k}}^K \!\! {{\mathrm{T}}_{ki}^{{\mathrm{p}}3}}  \!+\! {\sigma ^2}}} ,  \notag\\
  {\mathrm{SINR}}_k^{\mathrm{p}} &\!=\! \frac{{\frac{{{p_{\mathrm{d}}}}}{K}{\mathrm{T}}_{kk}^{{\mathrm{p}}1}}}{{\frac{{{p_{\mathrm{d}}}}}{K}\sum\limits_{i = 1}^K {{\mathrm{T}}_{ki}^{{\mathrm{p}}2}}  \!+\! \frac{{{p_{\mathrm{d}}}}}{K}\sum\limits_{i \in {\mathcal{P}_k}\backslash \left\{ k \right\}}^K {{\mathrm{T}}_{ki}^{{\mathrm{p}}1}}  \!+\! \frac{{{p_{\mathrm{d}}}}}{K}\sum\limits_{i \notin {\mathcal{P}_k}}^K {{\mathrm{T}}_{ki}^{{\mathrm{p}}3}}  \!+\! {\sigma ^2}}}, \notag
\end{align}
where each term is expressed as
\begin{align}
  {\mathrm{T}}_k^{{\mathrm{c}}1} &= {\left| {\sum\limits_{l = 1}^L {\sqrt {{\rho _l}{\mu _{{\mathrm{c}},l}}} \left( {\sum\limits_{i = 1}^K {{\mathbf{\bar h}}_{kl}^{\mathrm{H}}{{{\mathbf{\bar h}}}_{il}}}  + \sum\limits_{i \in {\mathcal{P}_k}}^K {{\mathrm{tr}}\left( {{{{\mathbf{\bar Q}}}_{kil}}} \right)} } \right)} } \right|^2} , \notag \\
  {\mathrm{T}}_k^{{\mathrm{c}}2} &=\! \sum\limits_{l = 1}^L {{\rho _l}} {\mu _{{\mathrm{c}},l}}\!\left( {\sum\limits_{i = 1}^K {\sum\limits_{j \in {\mathcal{P}_i}}^K {\left( {{\mathrm{tr}}\left( {{{{\mathbf{\bar Q}}}_{ijl}}{{\mathbf{R}}_{kl}}} \right) \!+\! {\mathbf{\bar h}}_{kl}^{\mathrm{H}}{{{\mathbf{\bar Q}}}_{ijl}}{{{\mathbf{\bar h}}}_{kl}}} \right)} } } \right.  \notag \\
  &\left. { + \sum\limits_{i = 1}^K {\sum\limits_{j = 1}^K {{\mathbf{\bar h}}_{il}^{\mathrm{H}}{{\mathbf{R}}_{kl}}{{{\mathbf{\bar h}}}_{jl}}} } } \right) , \notag \\
  {\mathrm{T}}_{ki}^{{\mathrm{p}}1} &= {\left| {\sum\limits_{l = 1}^L {\sqrt {\left( {1 - {\rho _l}} \right){\eta _{il}}{\mu _{il}}} \left( {{\mathbf{\bar h}}_{kl}^{\mathrm{H}}{{{\mathbf{\bar h}}}_{il}} + {\mathrm{tr}}\left( {{{{\mathbf{\bar Q}}}_{kil}}} \right)} \right)} } \right|^2} , \notag \\
  {\mathrm{T}}_{ki}^{{\mathrm{p}}2} &= \sum\limits_{i = 1}^K {\sum\limits_{l = 1}^L {\left( {1 \!-\! {\rho _l}} \right){\eta _{il}}{\mu _{il}}\left( {{\mathrm{tr}}\left( {{{\mathbf{Q}}_{il}}{{\mathbf{R}}_{kl}}} \right) \!+\! {\mathbf{\bar h}}_{kl}^{\mathrm{H}}{{\mathbf{Q}}_{il}}{{{\mathbf{\bar h}}}_{kl}}} \right.} }  \notag \\
  &\left. { + {\mathbf{\bar h}}_{il}^{\mathrm{H}}{{\mathbf{R}}_{kl}}{{{\mathbf{\bar h}}}_{il}}} \right) , \notag \\
  {\mathrm{T}}_{ki}^{{\mathrm{p}}3} &= {\left| {\sum\limits_{l = 1}^L {\sqrt {\left( {1 - {\rho _l}} \right){\eta _{il}}{\mu _{il}}} \left( {{\mathbf{\bar h}}_{kl}^{\mathrm{H}}{{{\mathbf{\bar h}}}_{il}}} \right)} } \right|^2} . \notag
\end{align}
Moreover, the normalized coefficients are derived as
\begin{align}
{\mu _{{\mathrm{c}},l}} &= {1}/\left({{\sum\limits_{k = 1}^K {\sum\limits_{i = 1}^K {{\mathbf{\bar h}}_{kl}^{\mathrm{H}}{{{\mathbf{\bar h}}}_{il}}}  + \sum\limits_{k = 1}^K {\sum\limits_{i \in {\mathcal{P}_k}}^K {{\mathrm{tr}}\left( {{{{\mathbf{\bar Q}}}_{kil}}} \right)} } } }}\right) , \notag\\
{\mu _{il}} &= {1}/\left({{{\mathbf{\bar h}}_{il}^{\mathrm{H}}{{{\mathbf{\bar h}}}_{il}} + {\mathrm{tr}}\left( {{{\mathbf{Q}}_{il}}} \right)}}\right) . \notag
\end{align}
\end{thm}
\begin{IEEEproof}
See Appendix A.
\end{IEEEproof}

Note that Theorem~\ref{thm:1} gives a general expression for CF massive MIMO systems. When RS is not considered, it reduces to the CF massive MIMO research under Rician channels discussed in \cite{8809413}. When the LoS path disappears, Theorem~\ref{thm:1} further simplifies to our previous work on RS-assisted CF massive MIMO under Rayleigh channels, as discussed in \cite{10032129}. In addition, when spatial correlation is absent, we obtain the following results.
\begin{cor}\label{cor1}
For spatially uncorrelated Rician fading, we can rewrite the variance of the channel and the estimate as ${{\mathbf{R}}_{kl}} \triangleq \beta _{kl}^{{\mathrm{nlos}}}{{\mathbf{I}}_N}$ and ${{\mathbf{Q}}_{kl}} \triangleq {\gamma _{kl}}{{\mathbf{I}}_N}$, where
\begin{align}
{\gamma _{kl}} = {p_k}{\tau _p}\beta _{kl}^{{\mathrm{nlos}}}{\Psi _{kl}}\beta _{kl}^{{\mathrm{nlos}}},
\end{align}
with
\begin{align}
{\Psi _{kl}} = {\left( {\sum\limits_{i \in {\mathcal{P}_k}} {{p_i}{\tau _p}\beta _{il}^{{\mathrm{nlos}}}}  + {\sigma ^2}} \right)^{ - 1}} .
\end{align}
Then, we can derive the lower bound on the sum SE, which is still represented by \eqref{eq:sumSE}, with ${{\mathrm{SINR}}_k^{\mathrm{c}}}$ and ${\mathrm{SINR}}_k^{\mathrm{p}}$, but it includes new terms as
\begin{align}
{\mathrm{T}}_k^{{\mathrm{c}}1} &= {\left| {\sum\limits_{l = 1}^L {\sqrt {{\rho _l}{\mu _{{\mathrm{c}},l}}} \left( {\sum\limits_{i = 1}^K {N\sqrt {\beta _{kl}^{{\mathrm{los}}}\beta _{il}^{{\mathrm{los}}}} }  + \sum\limits_{i \in {\mathcal{P}_k}}^K {N\sqrt {{\gamma _{kl}}{\gamma _{il}}} } } \right)} } \right|^2} ,\notag \\
{\mathrm{T}}_k^{{\mathrm{c}}2} &= \sum\limits_{l = 1}^L {{\rho _l}} {\mu _{{\mathrm{c}},l}}\!\left( {\sum\limits_{i = 1}^K {\sum\limits_{j \in {\mathcal{P}_i}}^K {\left( {N\beta _{kl}^{{\mathrm{nlos}}}\sqrt {{\gamma _{il}}{\gamma _{jl}}}  \!+\! N\beta _{kl}^{{\mathrm{los}}}\sqrt {{\gamma _{il}}{\gamma _{jl}}} } \right)} } } \right. \notag \\
& \left. { + \sum\limits_{i = 1}^K {\sum\limits_{j = 1}^K {N\beta _{kl}^{{\mathrm{nlos}}}\sqrt {\beta _{il}^{{\mathrm{los}}}\beta _{jl}^{{\mathrm{los}}}} } } } \right) , \notag \\
{\mathrm{T}}_{ki}^{{\mathrm{p}}1} &= {\left| {\sum\limits_{l = 1}^L {\sqrt {\left( {1 - {\rho _l}} \right){\eta _{il}}{\mu _{il}}} \left( {N\sqrt {\beta _{kl}^{{\mathrm{los}}}\beta _{il}^{{\mathrm{los}}}}  + N\sqrt {{\gamma _{kl}}{\gamma _{il}}} } \right)} } \right|^2} , \notag \\
{\mathrm{T}}_{ki}^{{\mathrm{p}}2} &= \sum\limits_{i = 1}^K {\sum\limits_{l = 1}^L {\left( {1 - {\rho _l}} \right){\eta _{il}}{\mu _{il}}\left( {N\beta _{kl}^{{\mathrm{nlos}}}{\gamma _{il}} + N\beta _{kl}^{{\mathrm{los}}}{\gamma _{il}}} \right.} }  \notag \\
& \left. { + N\beta _{il}^{{\mathrm{los}}}\beta _{kl}^{{\mathrm{nlos}}}} \right) , \notag \\
{\mathrm{T}}_{ki}^{{\mathrm{p}}3} &= {\left| {\sum\limits_{l = 1}^L {\sqrt {\left( {1 - {\rho _l}} \right){\eta _{il}}{\mu _{il}}} N\sqrt {\beta _{kl}^{{\mathrm{los}}}\beta _{il}^{{\mathrm{los}}}} } } \right|^2} , \notag
\end{align}
with the normalized coefficients including
\begin{align}
{\mu _{{\mathrm{c}},l}} &= 1/\left( {\sum\limits_{k = 1}^K {\sum\limits_{i = 1}^K {N\sqrt {\beta _{kl}^{{\mathrm{los}}}\beta _{il}^{{\mathrm{los}}}} }  + \sum\limits_{k = 1}^K {\sum\limits_{i \in {\mathcal{P}_k}}^K {N\sqrt {{\gamma _{kl}}{\gamma _{il}}} } } } } \right) , \notag \\
{\mu _{il}} &= 1/\left( {N\beta _{il}^{{\mathrm{los}}} + N{\gamma _{il}}} \right) . \notag
\end{align}
Significantly, assuming $N=1$ and without RS and the LoS path, we can derive \cite[Eq. (24)]{Ngo2017Cell} as a special case.
\end{cor}
\begin{rem}
Based on the expressions in Corollary \ref{cor1}, we focus only on the power splitting factor. When the number of antennas tends to infinity ($LN \to \infty $), the approximate sum SE can be derived as $\left( {\dot \rho /\left( {1 - \dot \rho } \right)} \right)\dot a + \dot b$. Here, $ \dot \rho $ refers to the power-splitting factor of the total system. Moreover, $\dot a$ is a positive constant reflecting the capacity of common messages. Meanwhile, $\dot b$ is another positive constant representing the capacity of private messages, which becomes fixed as the number of antennas approaches infinity. Therefore, an interesting insight can be found: when dealing with a CF massive MIMO system with a large number of antennas, it is necessary to increase the power-splitting factor as much as possible.
\end{rem}

\subsection{Power-Splitting Factors}

We first consider our previous work \cite{10032129} as a benchmark, which assumed equal power-splitting factors across all APs, thereby limiting the performance of RS-based CF massive MIMO systems. Then, we use the global statistical CSI from CF massive MIMO systems to design general and practical power-splitting factors for common messages. Inspired by the ``efficiency-driven'' concept\footnote{``Efficiency-driven'' is an economic concept that emphasizes enhancing competitiveness, finally leading to overall development of the system.}, we allocate larger power-splitting factors to APs with favorable CSI for total performance improvement,
resulting in our proposed heuristic algorithm where power-splitting factors are
\begin{align}
\rho_l &= \rho + \Delta_l , \forall l ,
\end{align}
where $0 \leqslant \rho \leqslant 1$ denotes the initial power-splitting factors, which are equal for all APs \cite{10032129}. Furthermore, $\Delta_l$ is the corresponding calibration coefficient, and it is given by
\begin{align}
{\Delta _l} = \omega \frac{{{\zeta _l} - \zeta }}{{\max \left\{ {\left| {{\zeta _l} - \zeta } \right|} \right\}}} ,
\end{align}
with $\omega  = \min \left\{ {\rho  - 0,1 - \rho } \right\}/\varepsilon$, ${\zeta _l} = {\left( {\left( {\sum\nolimits_{k = 1}^K {{\zeta _{kl}}} } \right)/K} \right)^a }$, $\zeta  = \left( {\sum\nolimits_{l = 1}^L {{\zeta _l}} } \right)/L $,
where $\omega$ denotes the calibration range of the power-splitting factor, $\zeta_l$ represents the average statistical CSI that reflects the capability of the $l$th AP to serve all UEs, and $\zeta$ is the average statistical CSI of the entire network representing a baseline of service capabilities. Moreover, $\varepsilon$ and $a$ are the scaling factors for the calibration range and the average statistical CSI, respectively, work well at values of $1.2$ and $1/2$.
The heuristic algorithm proposed above, which is based on statistical CSI, is computationally efficient with low complexity. Therefore, it is able to identify viable power-splitting factors, making it especially useful when a size of the CF massive MIMO system is large. However, it may not achieve an ideal solution, providing instead near-optimal alternatives.

To further approach upper bound performance of various power-splitting factors in the RS-based CF massive MIMO system, we define the problem of maximizing sum SE as follows:
\begin{align}\label{maxrho}
  \mathop {\max }\limits_{{\rho _l}}& {\text{ sum\;SE}}\left\{ {{\rho _l}} \right\} \notag \\
  {\text{s}}{\text{.t}}{\text{. }}&0 \leqslant {\rho _l} \leqslant 1,\forall l \hfill ,
\end{align}
where the sum SE is given by \eqref{eq:sumSE}. When the number of APs is not large, a genetic algorithm can be employed to derive a optimal solution for the optimization problem \eqref{maxrho}. Genetic algorithms are adaptive search methods based on natural selection and genetics \cite{katoch2021review}. They apply multiple solutions to a problem, scoring each based on its quality. The top solutions are then combined and slightly altered, creating new ones. Through repeated cycles, we gradually approach the optimal solution\footnote{It is necessary to save optimal solutions from all previous iterations at each selection operation, so that the genetic algorithm can eventually converge to the global optimum \cite{265964}.}.

\subsection{Power-Control Coefficients}

In wireless communication, power-control is the key to manage interference. It directly improves system capacity by maintaining robust signal quality with other UEs on the shared channel. Here, we propose a heuristic algorithm for power-control coefficients of private messages by utilizing global statistical CSI. Taking into account both the transceivers, we obtain
\begin{align}\label{eq:eta}
{\eta _{kl}} = \frac{{{\zeta _k}}}{{\max \left\{ {{\zeta _{k'}}} \right\}}} \times \frac{{\min \left\{ {{\zeta _{l'}}} \right\}}}{{{\zeta _l}}},
\end{align}
with ${\zeta _k} = {\left( {\left( {\sum\nolimits_{l' = 1}^L {{\zeta _{kl'}}} } \right)/L} \right)^{\bar a }}$, ${\zeta _l} = {\left( {\left( {\sum\nolimits_{k' = 1}^K {{\zeta _{k'l}}} } \right)/K} \right)^{\bar a } }$,
where ${\bar a} $ is the scaling factor for the average statistical CSI and performs well when the value is $1/4$.
It should be noted that in \eqref{eq:eta}, the ${{{\zeta _k}}}/{{\max \left\{ {{\zeta _{k'}}} \right\}}}$ of the expression suggests that UEs with better CSI are allocated more power to improve overall performance. On the other hand, the ${{\min \left\{ {{\zeta _{l'}}} \right\}}}/{{{\zeta _l}}}$ emphasizes that APs with poor service capabilities are given increased power to ensure network fairness. After this operation, the signal power of all UEs can be received at a superior level to increase the sum SE of the system.

Moreover, a power-control problem that maximizes the sum SE \eqref{eq:sumSE} can be formulated as\footnote{Here, we limit the maximum power per UE to avoid the possibility of a small number of UEs occupying power resources. The looser constraint $0 \leqslant \sum\nolimits_{i = 1}^K {{\eta _{il}}}  \leqslant K,\forall l$ wil be left for future work to improve performance.}
\begin{align}
  \mathop {\max }\limits_{{\eta _{il}}} & {\text{ sum\;SE}}\left\{ {{\eta _{il}}} \right\} \notag \\
  {\text{s}}{\text{.t}}{\text{. }} & 0 \leqslant {\eta _{il}} \leqslant 1,\forall i,l ,
\end{align}
which is a non-convex problem, and can also be solved by using the genetic algorithms to achieve the upper bound sum SE for comparison with proposed heuristic algorithms.
\begin{rem}
However, the genetic algorithms can take a long time to converge to a optimal solution, especially in complex scenarios. When a CF massive MIMO system has a large number of APs and UEs, employing the genetic algorithms to optimize the $L$ power-splitting factors and the $K \times L$ power-control coefficients demands significant computational resources, leading to high cost and time inefficiencies.
\end{rem}

\begin{figure*}[t]
\centering
\includegraphics[scale=0.45]{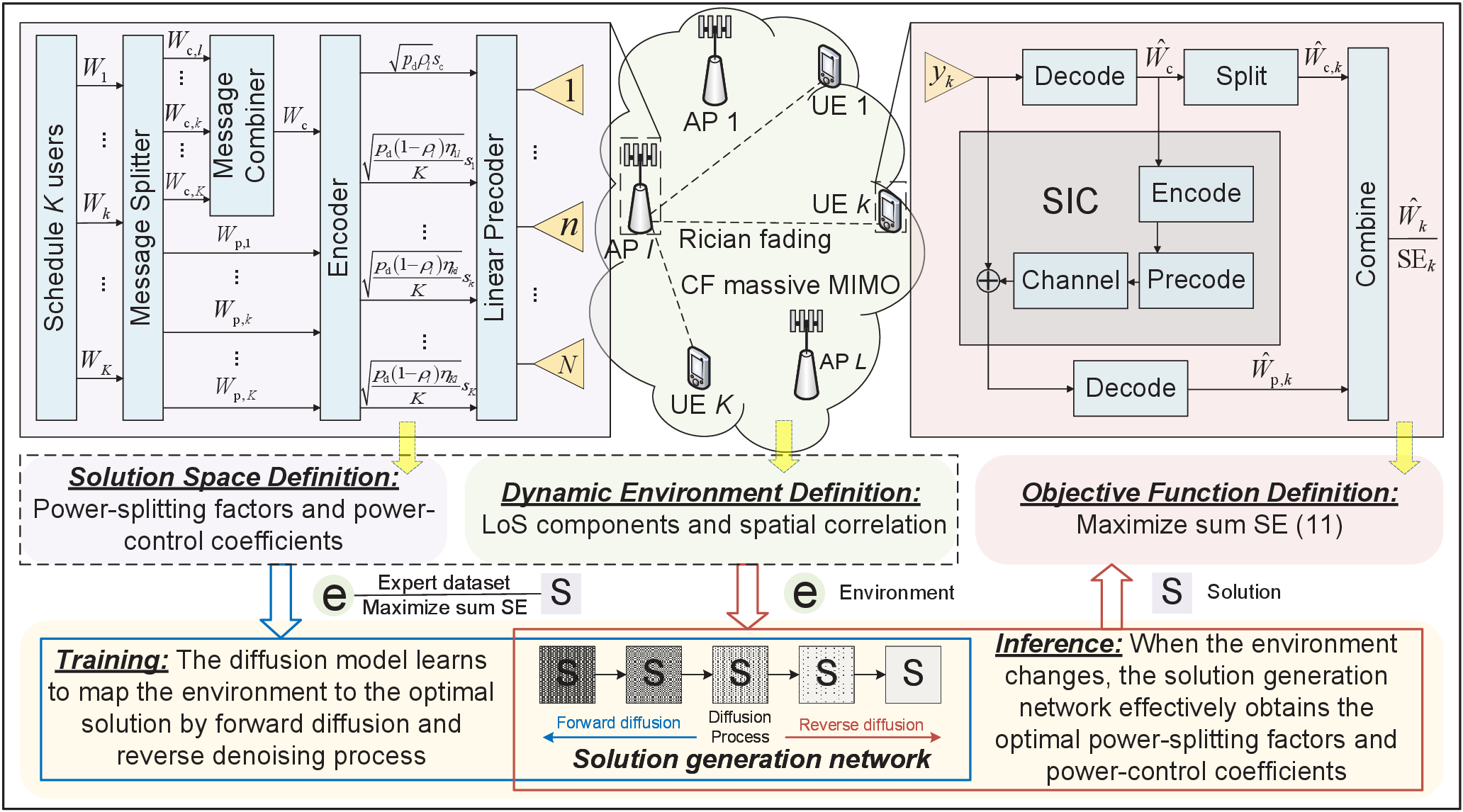}
\caption{GAI for the RS-based CF massive MIMO system under dynamic environments.} \vspace{-4mm}
\label{fig:CF_RS_GAI}
\end{figure*}

\section{Joint Optimization With GAI}\label{se:AIGC}

From the above discussion, it is clear that the optimization of power splitting factors and power-control coefficients is critical to achieve upper bound performance of the RS-based CF massive MIMO system.
The proposed heuristic algorithms based on statistical CSI offer efficient but suboptimal solutions. In contrast, the genetic algorithms approach an optimal solution, but at the cost of long convergence time.
Therefore, in this section, we jointly optimize the power-splitting factors and power-control coefficients by using a GAI approach \cite{du2023beyond}.

\subsection{Problem Formulation}

For a downlink, taking into account realizations of the large-scale fading, we optimize the power-splitting factors for APs ${\rho _l}$, for $l = 1, \ldots ,L$ and power-control coefficients for UEs ${\eta _{il}}$, for $i = 1, \ldots ,K$ and $l = 1, \ldots ,L$, aiming to maximize the sum SE \eqref{eq:sumSE} under the power constraint. Then, we have the following joint optimization problem:
\begin{align}\label{eq:problem}
  \mathop {\max }\limits_{{\rho _l},{\eta _{il}}} & {\text{ sum\;SE}}\left\{ {{\rho _l},{\eta _{il}}} \right\} \notag \\
  {\text{s}}{\text{.t}}{\text{. }} & 0 \leqslant {\rho _l} \leqslant 1,\forall l \notag \\
  & 0 \leqslant {\eta _{il}} \leqslant 1,\forall i,l ,
\end{align}
where the first constraint is to ensure that the transmit power of each AP can be allocated to both a common message and private message. The second constraint is to ensure the power limitation for private messages. Note that \eqref{eq:problem} is a non-convex problem with totally $L + K \times L$ variables. It is challenging to solve with standard methods, often entailing long convergence time to obtain a solution, particularly in large-scale networks. Moreover, in dynamic wireless environments, time efficiency of the solution determines its practical usefulness. Hence, considering variable LoS components and spatial correlation in the Rician channel, we propose a GAI-based algorithm by using a diffusion model to obtain power-splitting factors and power-control coefficients.

\subsection{Generative AI Approach}

The basic principle of the GAI approach is simple as illustrated in Fig.~\ref{fig:CF_RS_GAI}.
We leverage the diffusion model to obtain solutions of this optimization problem \eqref{eq:problem}. The major two processes include:
\subsubsection{Forward diffusion}

The forward diffusion process is modeled as a Markov chain, sequentially adding noise\footnote{Note that the noise here is not the wireless noise in the previous system but the Gaussian noise added in the diffusion model.} to power-splitting factors and power-control coefficients over $T$ steps. For a given expert solution ${\mathbf{x}}_0 = \left\{ {{\rho _l},\forall l;{\eta _{il}},\forall i,l} \right\}$, this forward process can be expressed as
\begin{align} \label{eq:x1T}
  f\left( {{{\mathbf{x}}_{1:T}}\left| {{{\mathbf{x}}_0}} \right.} \right) &= \prod\limits_{t = 1}^T {f\left( {{{\mathbf{x}}_t}\left| {{{\mathbf{x}}_{t - 1}}} \right.} \right)}  \notag \\
   &\!\!\!\!\!\!\!\!\!= \prod\limits_{t = 1}^T {\mathcal{N}\left( {{{\mathbf{x}}_t};{\mathbf{ m }_t} = \sqrt {1 - {v_t}} {{\mathbf{x}}_{t - 1}},{{\mathbf{V}}_t} = {v_t}{\mathbf{I}}} \right)} ,
\end{align}
where ${f\left( {{{\mathbf{x}}_t}\left| {{{\mathbf{x}}_{t - 1}}} \right.} \right)}$ follows a Gaussian distribution defined by the mean ${\mathbf{ m }_t}$ and the variance ${{\mathbf{V}}_t}$, with ${\mathbf{I}}$ being the identity matrix signifying a uniform standard deviation $v_t$ across all dimensions. It is clear from \eqref{eq:x1T} that when $t$ is large, deriving ${{\mathbf{x}}_t}$ requires a large number of computational steps. To reduce the computational time cost, we define ${\alpha _t} = 1 - {v_t}$ and ${{\bar \alpha }_t} = \prod\nolimits_{i = 1}^t {\left( {1 - {v_i}} \right)} $, which facilitates the following simplifications:
\begin{align}
  {{\mathbf{x}}_t} &= \sqrt {{\alpha _t}} {{\mathbf{x}}_{t - 1}} + \sqrt {1 - {\alpha _t}} {\bm{\epsilon}_{t - 1}} = \sqrt {{\alpha _t}{\alpha _{t - 1}}} {{\mathbf{x}}_{t - 2}} \notag\\
  &+ \sqrt {1 - {\alpha _t}{\alpha _{t - 1}}} {\bm{\epsilon}_{t - 2}} =  \ldots  = \sqrt {{{\bar \alpha }_t}} {{\mathbf{x}}_0} + \sqrt {1 - {{\bar \alpha }_t}} {\bm{\epsilon}_0} ,
\end{align}
where the added Gaussian noise $\bm{\epsilon}_{i} \sim \mathcal{N}\left( {{\mathbf{0}},{\mathbf{I}}} \right), \forall i$. Then, ${{\mathbf{x}}_t}$ can be derived from the distribution given below:
\begin{align}
{{\mathbf{x}}_t} \sim f\left( {{{\mathbf{x}}_t}\left| {{{\mathbf{x}}_0}} \right.} \right) = \mathcal{N}\left( {{{\mathbf{x}}_t};\sqrt {{{\bar \alpha }_t}} {{\mathbf{x}}_0},\left( {1 - {{\bar \alpha }_t}} \right){\mathbf{I}}} \right) .
\end{align}
Given that ${v_t}$ is a hyperparameter, we can precompute ${\alpha _t}$ and ${{\bar \alpha }_t}$ at all timesteps. Note that the variance parameter ${v_t}$ may either be set as a constant or selected according to a ${v_t}$-schedule across $T$ timesteps \cite{ho2020denoising}.

\subsubsection{Reverse diffusion}

As $T$ approaches infinity, power-splitting factors and power-control coefficients converges to a standard Gaussian noise distribution ${{\mathbf{x}}_T} \sim \mathcal{N}\left( {{\mathbf{0}},{\mathbf{I}}} \right)$. Theoretically, if ${f\left( {{{\mathbf{x}}_{t-1}}\left| {{{\mathbf{x}}_{t}}} \right.} \right)}$ can be obtained, we can use it to recover the original power-splitting factors and power-control coefficients from the standard Gaussian distribution. However, estimation of ${f\left( {{{\mathbf{x}}_{t-1}}\left| {{{\mathbf{x}}_{t}}} \right.} \right)}$ necessitates knowledge of all power-splitting and power-control strategies in all conditions, which is practically intractable. Therefore, a neural network is used to learn the following transition relation:
\begin{align}
{r_\theta }\left( {{{\mathbf{x}}_{t - 1}}\left| {{{\mathbf{x}}_t}} \right.} \right) = \mathcal{N}\left( {{{\mathbf{x}}_{t - 1}};{{\mathbf{m }}_\theta }\left( {{{\mathbf{x}}_t},t} \right),{{\mathbf{V }}_\theta }\left( {{{\mathbf{x}}_t},t} \right)} \right) ,
\end{align}
where $\theta $ the hyperparameter of the neural network. On this basis, the inverse denoising process is described as
\begin{align}
  {r_\theta }\left( {{{\mathbf{x}}_{0:T}}} \right) &= {r_\theta }\left( {{{\mathbf{x}}_T}} \right)\prod\limits_{t = 1}^T {{r_\theta }\left( {{{\mathbf{x}}_{t - 1}}\left| {{{\mathbf{x}}_t}} \right.} \right)}  \notag \\
   &\!\!\!\!\!\!= {r_\theta }\left( {{{\mathbf{x}}_T}} \right)\prod\limits_{t = 1}^T {\mathcal{N}\left( {{{\mathbf{x}}_{t - 1}};{{\mathbf{m}}_\theta }\left( {{{\mathbf{x}}_t},t} \right),{{\mathbf{V}}_\theta }\left( {{{\mathbf{x}}_t},t} \right)} \right)} .
\end{align}
According to \cite{ho2020denoising}, ${{r_\theta }\left( {{{\mathbf{x}}_{t - 1}}\left| {{{\mathbf{x}}_t}} \right.} \right)}$ in the denoising process can be regarded as a noise prediction model where the covariance matrix fixed as $ {{\mathbf{V}}_\theta }\left( {{{\mathbf{x}}_t},t} \right) = {v_t}{\mathbf{I}}$, with the mean calculated as
\begin{align}
{{\mathbf{m}}_\theta }\left( {{{\mathbf{x}}_t},t} \right) = \frac{1}{{\sqrt {{\alpha _t}} }}\left( {{{\mathbf{x}}_t} - \frac{{{v_t}}}{{\sqrt {1 - {{\bar \alpha }_t}} }}{\bm{\epsilon}_\theta }\left( {{{\mathbf{x}}_t},t} \right)} \right) .
\end{align}
We start by sampling ${{\mathbf{x}}_T} \sim \mathcal{N}\left( {{\mathbf{0}},{\mathbf{I}}} \right)$ and then use parameter $\theta$ to form the reverse diffusion chain as
\begin{align}\label{eq:reverse}
{{\mathbf{x}}_{t - 1}}\left| {{{\mathbf{x}}_t}} \right. = \frac{{{{\mathbf{x}}_t}}}{{\sqrt {{\alpha _t}} }} - \frac{{{v_t}}}{{\sqrt {{\alpha _t}\left( {1 - {{\bar \alpha }_t}} \right)} }}{\bm{\epsilon}_\theta }\left( {{{\mathbf{x}}_t},t} \right) + \sqrt {{v_t}}\bm{\epsilon} ,
\end{align}
where $\bm{\epsilon} \sim \mathcal{N}\left( {{\mathbf{0}},{\mathbf{I}}} \right)$. Furthermore, the authors in \cite{ho2020denoising} introduced simplifications to the original loss function by disregarding a specific weighting term:
\begin{align}\label{eq:loss}
{\mathcal{L}_t} = {\mathbb{E}_{{{\mathbf{x}}_0},t,\bm{\epsilon}}}\left\{ {{{\left\| {\bm{\epsilon} - {\bm{\epsilon}_\theta }\left( {\sqrt {{{\bar \alpha }_t}} {{\mathbf{x}}_0} + \sqrt {1 - {{\bar \alpha }_t}}\bm{\epsilon} ,t} \right)} \right\|}^2}} \right\} .
\end{align}
This demonstrates that the model is designed to predict the noise $\bm{\epsilon}$ at each timestep, rather than the mean of the distribution.

\subsubsection{GAI as solution}

Subsequently, we detail the method for resolving the problem by applying the GAI approach. The diffusion model is trained to obtain power-splitting factors and power-control coefficients to maximize the sum SE, with the solution steps detailed as follows:
\begin{itemize}
\item \textit{\textbf{Solution Space Definition:}} The diffusion model accurately generates the power-splitting factors ${\rho _l},\forall l$ and power-control coefficients ${\eta _{il}},\forall i,\forall l$ through a sequence of denoising operations on Gaussian noise, guaranteeing the maximization of the sum SE. Therefore, the dimension of the solution vector is decided by the number of UEs and APs in the system.
\item \textit{\textbf{Objective Function Definition:}} Next, we define the objective function for the optimization problem, which could be either maximization or minimization, depending on the requirements of the network. The training objective in our study is to maximize the sum SE using power-splitting factors and power-control coefficients generated by the diffusion model. Furthermore, the time-consuming genetic algorithm serves as a valuable tool for establishing the upper bound, guiding the optimization towards optimal solutions.
\item \textit{\textbf{Dynamic Environment Definition:}} The wireless channel propagation environment is inherently dynamic and diverse, constantly influenced by a myriad of factors. To enhance practicality, the diffusion model is designed to generate optimal power-splitting factors and power-control coefficients under varying conditions. In the correlated Rician channels under consideration, we assume variability in both the LoS components and spatial correlations \cite{bjornson2017massive}. For instance, the Rician factor ${\bar K}$ may randomly fluctuate between $-10$ dB and $10$ dB, while the ASD may vary between ${10^\circ }$ and ${90^\circ }$\footnote{Both the Rician factor and ASD can be efficiently measured by professional equipment \cite{9237116}. Thus, we define them to represent the dynamic environment to obtain preliminary results. In the future work, ray-tracing and channel knowledge map technologies will be used to accurately characterize complex environments \cite{zeng2024tutorial}, enhancing data quality and improving system performance.}. Note that the upper and lower bounds of the changes can be adjusted flexibly according to specific requirements.
\item \textit{\textbf{Training and Inference:}} The conditional diffusion model is proposed to generate the solution. We denote the power-splitting and power-control solutions as $\mathbf{s}$, and the corresponding environment is $\mathbf{e}$. The diffusion model mapping environments to solutions refers to the solution generation network, i.e., ${\bm{\epsilon}_\theta }\left( {\mathbf{s}\left| \mathbf{e} \right.} \right)$ with $\theta$ is the neural network parameter. The purpose of ${\bm{\epsilon}_\theta }\left( {\mathbf{s}\left| \mathbf{e} \right.} \right)$ is to generate deterministic power-splitting factors and power-control coefficients in the given environment that maximizes the sum SE. Moreover, the solution generation network is depicted through the reverse diffusion process expressed in \eqref{eq:reverse}. The next step is to train the ${\bm{\epsilon}_\theta }$ network. With a dataset of expert solutions already available (e.g., obtaining by the genetic algorithm), the loss function aims to minimize the gap between the expert and generated sum SE, formulated as:
    \begin{align}
   \mathop {\arg \min }\limits_{{\bm{\epsilon}_\theta }} \mathcal{L}\left( \theta  \right) &= {\mathbb{E}_{{{\mathbf{s}}_0} \sim {\bm{\epsilon}_\theta }}}\left\{ {\left\| {R\left( {\left\{ {\bar K,{\text{ASD}}} \right\},{{\mathbf{s}}_0}} \right)} \right.} \right. \notag \\
  &\left. {{{\left. { - {R_{\exp }}\left( {\left\{ {\bar K,{\text{ASD}}} \right\}} \right)} \right\|}^2}} \right\} ,
    \end{align}
    where ${R}$ represents the sum SE value for the generated power-splitting and power-control solutions ${{\mathbf{s}}_0}$ in the environment $ \left\{ {\bar K,{\text{ASD}}} \right\}$, and ${{R_{\exp }}}$ represents the sum SE value of the expert solution under the same environment. Let ${{{\mathbf{x}}_0}}$ denote the expert solution of ${{R_{\exp }}}$ and with the help of \eqref{eq:loss}, the simplified loss function is expressed as
    \begin{align}\label{eq:losssimple}
    \mathop {\arg \min }\limits_{{\bm{\epsilon}_\theta }} & \mathcal{L}\left( \theta  \right) = \mathbb{E}\left\{ {\left\| \bm{\epsilon}  \right.} \right. \notag \\
    &\!\!\!\!\!\!\!\!\! \left. {{{\left. { - {\bm{\epsilon}_\theta }\!\left( {\sqrt {{{\bar \alpha }_t}} {{\mathbf{x}}_0} \!+\! \sqrt {1 \!-\! {{\bar \alpha }_t}} \bm{\epsilon},t,\left\{ {\bar K,{\text{ASD}}} \right\}} \right)} \right\|}^2}} \right\} ,
    \end{align}
    where $\bm{\epsilon}$ represents the added Gaussian noise, $\sqrt {{{\bar \alpha }_t}} {{\mathbf{x}}_0} + \sqrt {1 - {{\bar \alpha }_t}} \bm{\epsilon}$ is the expert solution after the forward diffusion process. Once the training process is complete, the solution generation network ${\bm{\epsilon}_\theta }$ is well-equipped to efficiently generate the optimal power-splitting factors and power-control coefficients according to \eqref{eq:reverse}, adapting to any changes in ${\bar K}$ and ASD. The detailed steps are presented in Algorithm \ref{GAI}.
\end{itemize}

Note that the denoising step is $T=10$, the exploration noise is $\bm{\epsilon}=0.01$, and the learning rate for the ${\bm{\epsilon}_\theta }$ network is $10^{-4}$.
We run our proposed algorithms on a Windows 11 system with 12th Gen Intel(R) Core(TM) i7-12700H CPU and NVIDIA GeForce RTX 3060 GPU. 
We compute a small number of expert solutions for training and inference, where Rician ${\bar K}$ takes values from $-10$ dB to $20$ dB with an interval of $1$ dB, and ASD takes values from $5^\text{o}$ to $90^\text{o}$ with an interval of $5^\text{o}$.

\renewcommand{\algorithmicrequire}{\textbf{Training Part:}}
\renewcommand{\algorithmicensure}{\textbf{Inference Part:}}
\begin{algorithm}[htb]
\caption{GAI-based Algorithm for Solving \eqref{eq:problem}}
\label{GAI}
\begin{algorithmic}[1]
\Require
\State Input parameters: denoising step $T$, exploration noise $\bm{\epsilon}$, initialize neural network parameters $\theta$;
\State \#\#\ \ \ \ \emph{Start Learning Process}
\State Initialize a random process $\mathcal{N}$ for power-splitting and power-control exploration;
\While{not converge}
\State Observe the current environment $ \mathbf{e} = \left\{ {\bar K,{\text{ASD}}} \right\}$;
\State Obtain expert solutions ${{{\mathbf{x}}_0}}$ based on the current environment, such as with the help of genetic algorithms;
\State Randomly generate Gaussian noise $\bm{\epsilon}$, then add noise to disrupt the expert solution ${{{\mathbf{x}}_T}} = \sqrt {{{\bar \alpha }_T}} {{\mathbf{x}}_0} + \sqrt {1 - {{\bar \alpha }_T}} \bm{\epsilon}$;
\State Generate the power-splitting and power-control solution ${{\mathbf{s}}_0}$ by applying ${\bm{\epsilon}_\theta }$ to denoise ${{\mathbf{x}}_T}$ using \eqref{eq:reverse};
\State Update ${\bm{\epsilon}_\theta }$ according to \eqref{eq:losssimple};
\EndWhile
\State \textbf{return} The trained solution generation network ${\bm{\epsilon}_\theta }$;
\end{algorithmic}
\begin{algorithmic}[1]
\Ensure
\State Get the environment vector $\mathbf{e}$;
\State Obtain the optimal power-splitting factor and power-control coefficient vector ${{\mathbf{s}}_0}$ by using ${\bm{\epsilon}_\theta }$ to denoise Gaussian noise;
\State \textbf{return} The optimal power-splitting and power-control vector ${{\mathbf{s}}_0}$;
\end{algorithmic}
\end{algorithm}

\subsubsection{Complexity Analysis}

The complexity of our GAI-based algorithm can be expressed as $\mathcal{O} (N_\text{gai}(G + Tw_a + w_c))$, where $N_\text{gai}$ is the number of iterations until convergence, $G$ is the complexity of the genetic algorithm for expert solution generation (which can be prepared in advance and thus ignored), $T$ is the denoising step as defined in the input parameters, and $w_a$ and $w_c$ are the actor and critic networks' weight counts, respectively \cite{10409284}. In comparison, the DRL-based algorithm often exhibits a complexity of $\mathcal{O}(N_\text{drl}(G + w_a + w_c))$ per episode, where $N_\text{drl}$ is the number of iterations.

The complexity of our method relies more on the convergence rate and the denoising process, potentially enhancing computational intensity. However, this increased complexity translates to enhanced exploration capabilities in complex power-splitting and power-control scenarios, particularly in environments that the DRL-based algorithm might struggle to efficiently navigate high-dimensional state and action spaces. Importantly, while our approach may appear more computationally demanding, it often achieves comparable performance with fewer iterations. In many scenarios, $N_\text{gai}$ can be smaller than $N_\text{drl}$ to reach similar performance levels, effectively mitigating the additional per-iteration complexity \cite{du2023beyond}.

\section{Numerical Results and Discussion}\label{se:numerical}

Utilizing the three-slope propagation model \cite{Ngo2017Cell,10032129}, we consider a simulation environment where $L$ APs and $K$ UEs are uniformly and independently distributed within a square of size $500$ m $\times$ $500$ m.
The system operates at a carrier frequency of $f_c=2$ GHz, with a bandwidth allocation of $B=20$ MHz, and the noise power is set at $\sigma^2=-96$ dBm \cite{bjornson2019making}. Moreover, the transmission power for the pilot is set at $p = 20$ dBm, while that for the data is $p_\text{d} = 23$ dBm. Furthermore, a coherence block has a length of $T_c = 2$ ms and includes $\tau_c = 200$ channel uses \cite{bjornson2017massive}.

\begin{figure}[t]
\centering
\includegraphics[scale=0.6]{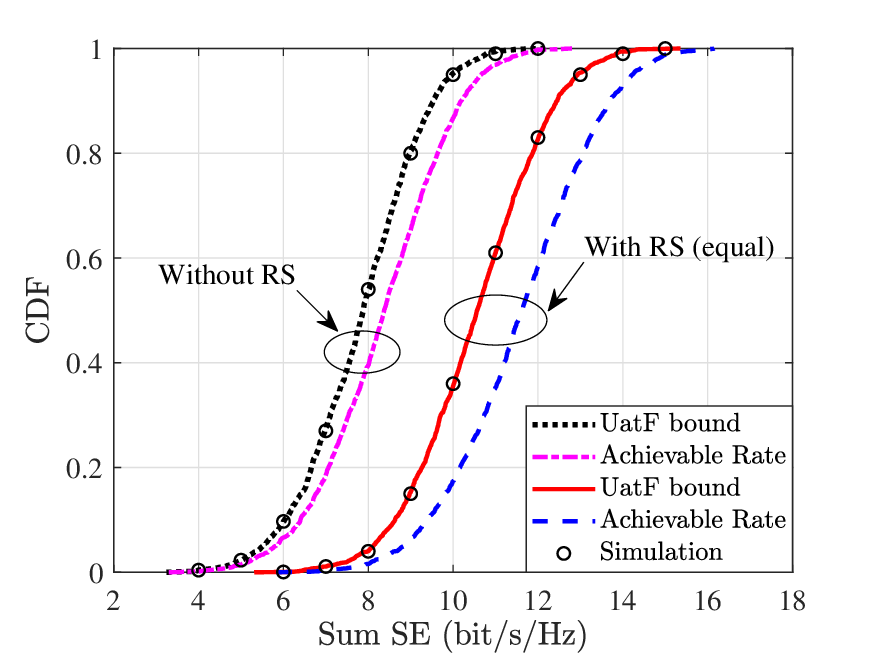}
\caption{CDF of the sum SE for RS-based CF massive MIMO systems ($K=4$, $L=20$, $N=4$, $\bar K = 5$ dB, $\tau_p=K/2$).} \vspace{-4mm}
\label{fig:CDF}
\end{figure}

\begin{figure}[t]
\centering
\includegraphics[scale=0.6]{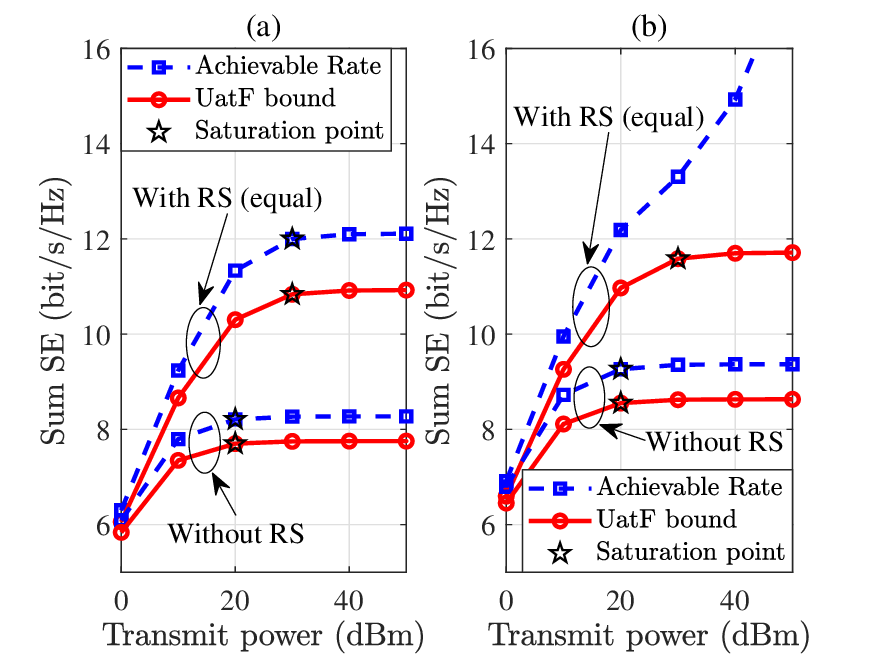}
\caption{Sum SE for RS-based CF massive MIMO systems against transmit power per AP ($K=4$, $L=20$, $N=4$, $\bar K = 5$ dB). (a) Imperfect CSI with pilot contamination and noise interference; (b) Perfect CSI without pilot contamination and noise interference.} \vspace{-4mm}
\label{fig:power}
\end{figure}

Fig.~\ref{fig:CDF} presents the CDF of the sum SE for the RS-based CF massive MIMO systems, highlighting a comparison between the achievable rate and the UatF bound. It is clear that RS significantly enhances the sum SE, both for the achievable rate and the UatF bound, with median sum SE gains of $3.2$ bit/s/Hz and $2.7$ bit/s/Hz, respectively. This is mainly due to the ability of RS to reduce the interference levels in the CF massive MIMO network through the application of successive interference cancellation. Furthermore, the median sum SE differences between the achievable rate and the UatF bound, with and without RS, are $1$ bit/s/Hz and $0.5$ bit/s/Hz, respectively. It is worth noting that the case without RS refers to the conventional CF massive MIMO, which is equivalent to the research in \cite{8809413} under the Rician channel. Moreover, the differences in the average SE of UEs between the UatF bound and the achievable rate do not exceed $1/4$ bit/s/Hz. Therefore, the closed-form expression for sum SE based on the UatF bound that we have derived serves as a valuable and tight lower bound, as the high antenna density ($320 \text{ antennas}/\text{km}^2$) enhances the channel hardening \cite{8379438}. 
Furthermore, simulation results have validated the accuracy of the derived closed-form expressions.
Note that the derived expressions work for any number of UEs, and using fewer UEs can accelerate the simulation process. In the future work, it will be necessary to enhance RS interference suppression capabilities to serve more UEs in extreme interference environments.

Taking into account both imperfect and perfect CSI, which corresponds to cases with and without pilot contamination and noise interference during channel estimation, Fig.~\ref{fig:power} shows the sum SE for the RS-based CF massive MIMO systems against transmit power per AP. In the case of imperfect CSI, illustrated in Fig.~\ref{fig:power} (a), it is obvious that the sum SE gradually approaches a saturation level with increasing transmit power, and this trend is observed both with and without RS.
This is because an increase in the transmission power of the desired signal simultaneously amplifies the multi-user interference and the interference resulting from imperfect CSI. It is commendable that RS can remove part of the interference leading to delayed saturation, so it can continue to increase transmit power to achieve performance gain. For the case of perfect CSI in Fig.~\ref{fig:power} (b), it is observed that the sum SE without RS is still saturated due to multi-user interference. Fortunately, due to the ability of RS to decode multi-user interference, the sum SE based on achievable rate will always increase as the transmit power increases.
However, the sum SE based on the UatF bound remains saturated due to the residual interference caused by the beamforming gain uncertainty term in \eqref{eq:r_c}.

\begin{figure}[t]
\centering
\includegraphics[scale=0.6]{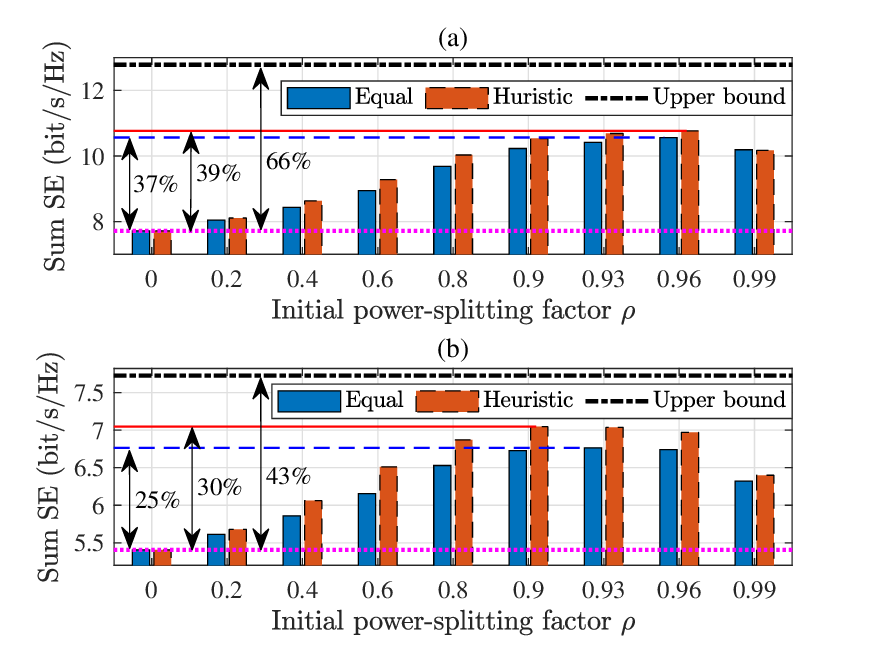}
\caption{Sum SE for RS-based CF massive MIMO systems against initial power-splitting factor with different power-splitting algorithms ($K\!=\!4$, $L\!=\!20$, $N\!=\!4$, $\tau_p\!=\!K/2$). (a) Rician channel with $\bar K \!=\! 5$ dB; (b) Rayleigh channel.} \vspace{-4mm}
\label{fig:power-splitting}
\end{figure}

\begin{figure}[t]
\centering
\includegraphics[scale=0.6]{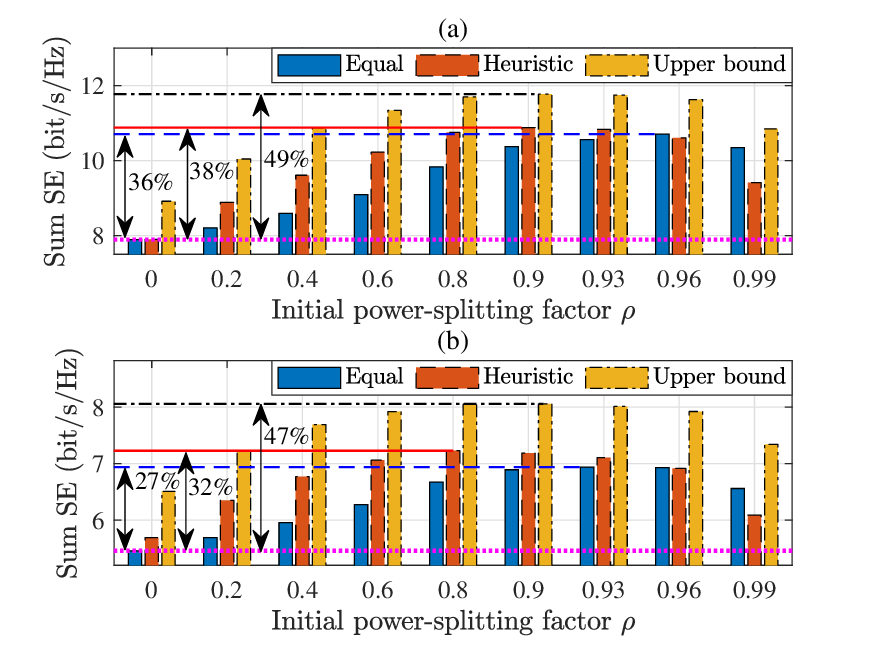}
\caption{Sum SE for RS-based CF massive MIMO systems against initial power-splitting factor with different power-control algorithms ($K=4$, $L=20$, $N=4$, $\bar K = 5$ dB, $\tau_p=K/2$). (a) Rician channel with $\bar K \!=\! 5$ dB; (b) Rayleigh channel.} \vspace{-4mm}
\label{fig:power-control}
\end{figure}

The sum SE for the RS-based CF massive MIMO systems against initial power-splitting factor $\rho$ with different power-splitting algorithms is illustrated in Fig.~\ref{fig:power-splitting}.
It is observed that the sum SE first increases and then decreases as the initial power-splitting factor $\rho$ increases, indicating that there is an optimal balance between power allocated to common and private messages for achieving peak performance. 
Additionally, achieving optimal sum SE performance in the Rician channel requires a larger $\rho$ than that in the Rayleigh channel. This is because the LoS components amplify the interference levels in the network, necessitating a greater allocation of power to common messages to decode the interference effectively\footnote{Therefore, increasing the network density, the number of UEs, and the transmit power, as well as other operations that increase interference, drive us to split the message into common and private parts. The larger the power-splitting factor, the better the performance.}.
Moreover, the proposed heuristic algorithm performs better in the Rayleigh channel compared with that in the Rician channel. For example, in the Rician channel, the equal power-splitting, the heuristic power-splitting, and the genetic algorithm-based upper bound schemes show improvements of 37\%, 39\%, and 66\%, respectively, compared with the case without RS. Moreover, the corresponding improvements in the Rayleigh channel are 25\%, 30\%, and 43\%, respectively.

Moreover, similar methods are also applied to the design of power-control to obtain the equal power-control algorithm, the heuristic power-control algorithm, and the genetic-based power-control algorithm \cite{katoch2021review}, respectively, as illustrated in Fig.~\ref{fig:power-control}. 
Interestingly, implementing effective power-control can also lead to a reduction in the initial power-splitting factor $\rho$ at the point where we obtain maximum sum SE. This is because power-control effectively mitigates interference, thereby reducing the reliance on RS for interference decoding\footnote{Therefore, when using efficient interference suppression algorithms such as power control and beamforming, the burden of RS handling interference is shared. A small power-splitting factor is sufficient to achieve optimal performance.}.
Moreover, the heuristic power-control algorithm shows enhanced performance in the Rayleigh channel when compared with that in the Rician channel.
Despite the exceptional performance of genetic algorithms, their considerable computational requirements limit their practicality. Hence, there is a pressing need for more advanced solutions that can meet the obtained upper bound.

\begin{figure}[t]
\centering
\includegraphics[scale=0.6]{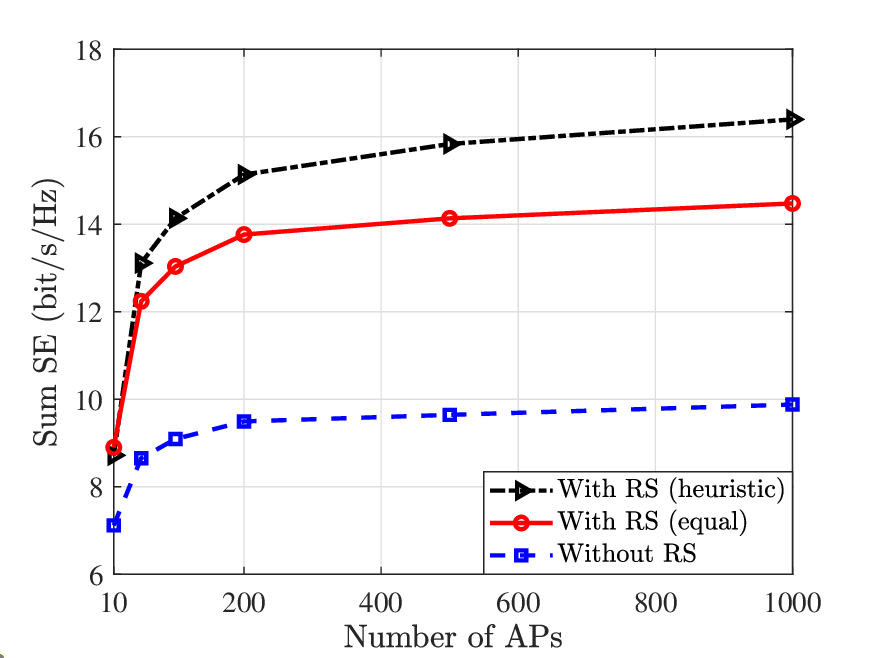}
\caption{Sum SE for RS-based CF massive MIMO systems against the number of APs ($K=4$, $N=4$, $\bar K = 5$ dB, $\tau_p=K/2$).} \vspace{-4mm}
\label{fig:NumOfAP}
\end{figure}

\begin{figure}[t]
\centering
\includegraphics[scale=0.6]{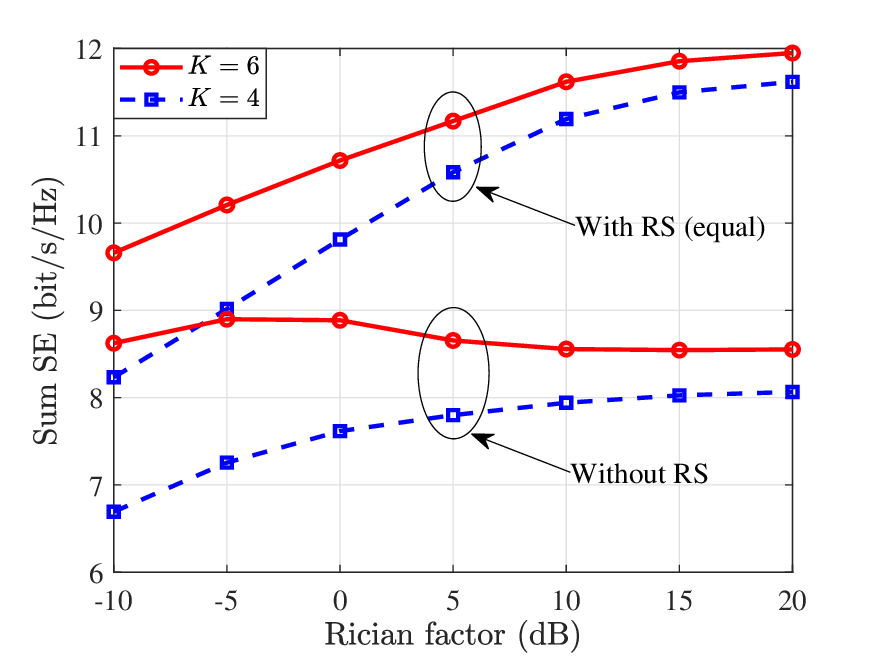}
\caption{Sum SE for RS-based CF massive MIMO systems against Rician factors ($K=4$, $L=20$, $N=4$, $\tau_p=K/2$).} \vspace{-4mm}
\label{fig:RicianFactor}
\end{figure}

Fig.~\ref{fig:NumOfAP} compares the sum SE for RS-based CF massive MIMO systems against the number of APs, evaluating three solutions: (i) without RS, (ii) with RS, and (iii) with heuristic enhanced RS. Note that heuristic enhanced RS refers to  the utilization of statistical CSI for the joint optimization of power-splitting factors and power-control coefficients. Clearly, as the number of APs increases, the rate of increase in the sum SE gradually slows down, especially when RS is not used. For example, as the number of APs grows from $100$ to $500$, the sum SE improvements for the corresponding three solutions are $0.5$ bit/s/Hz, $1.1$ bit/s/Hz, and $1.7$ bit/s/Hz, respectively. The reason is that increasing the number of APs brings both antenna gain and interference within the network, while RS can decode this interference, and our proposed solution is even more effective at mitigating it.

Fig.~\ref{fig:RicianFactor} shows the sum SE for RS-based CF massive MIMO systems against Rician factors, comparing cases with and without RS. 
In the case without RS and with $4$ UEs, it is observed that the sum SE increases with the Rician factor, but at a diminishing rate. This happens because while the LoS path strengthens the desired signal, it also increases the possibility of interference. For instance, when RS is not employed and the number of UEs is set to 6, it is observed that as the Rician factor exceeds $0$ dB, inter-user interference becomes the dominant factor, resulting in the sum SE no longer increasing as the Rician factor increases. Fortunately, RS mitigates interference, leading to a consistent increase in the sum SE. 
It is worth noting that with more UEs, the resulting higher interference may lead to a performance decrease in the RS case as the Rician factor increases. However, RS can always delay the performance degradation compared to the case without RS.
Another interesting observation is that in LoS-dominant settings, adding more UEs barely enhances the sum SE due to the reduced spatial multiplexing caused by the correlated LoS paths.

\begin{figure}[t]
\centering
\includegraphics[scale=0.6]{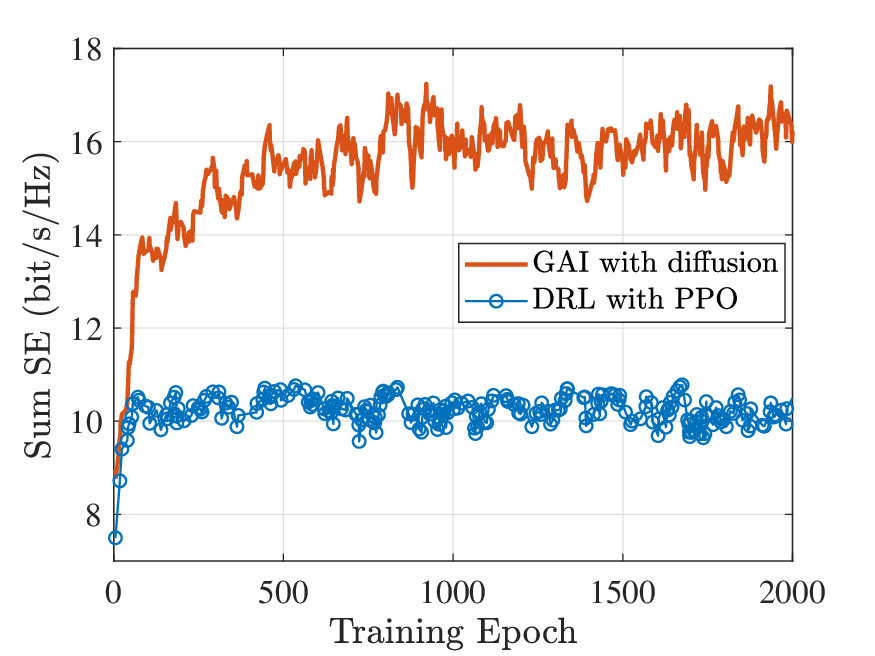}
\caption{Sum SE curves of GAI-based and DRL-based algorithms ($K=4$, $L=20$, $N=4$, $\tau_p=K/2$).} \vspace{-4mm}
\label{fig:coverage}
\end{figure}
%; The training process with the diffusion step of $10$, batch size of $512$, soft target update parameter of $0.005$, discount factor of $0.95$, exploration noise of $0.01$, and learning rate of $10^{-4}$

\begin{figure}[t]
\centering
\includegraphics[scale=0.6]{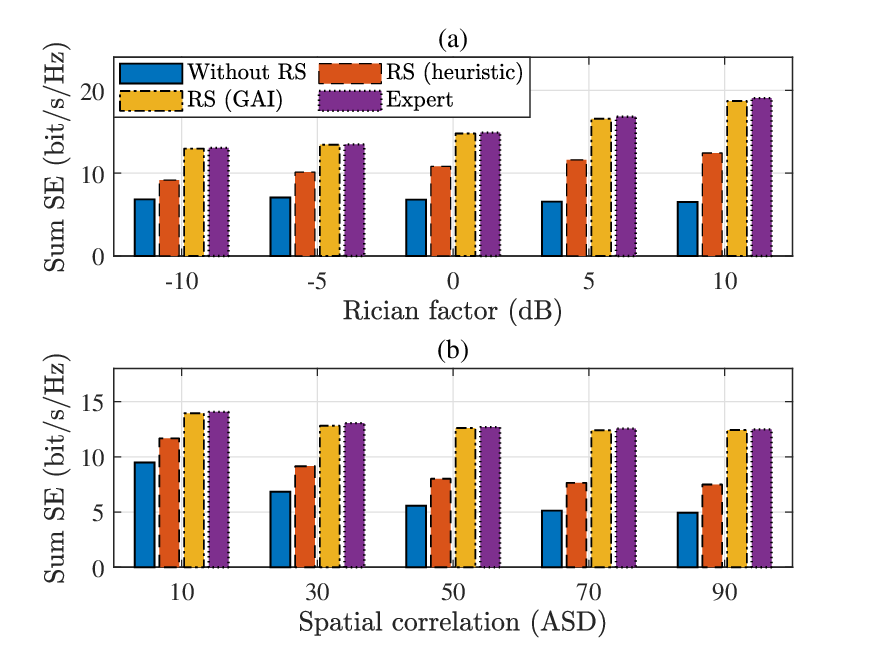}
\caption{Sum SE of the GAI-based algorithm under dynamic environments ($K=4$, $L=20$, $N=4$, $\tau_p=K/2$).} \vspace{-4mm}
\label{fig:GAI}
\end{figure}

Fig.~\ref{fig:coverage} illustrates the sum SE against the training epoch for the proposed GAI-based algorithm, in comparison with the DRL-based algorithm that uses Proximal Policy Optimization (PPO) \cite{10158526}. 
To ensure a fair comparison, both the DRL and GAI approaches were trained and tested under identical conditions, datasets, and common hyperparameters, with extensively tuned unique hyperparameters for optimal performance \cite{10409284}.
It can be observed that the DRL-based algorithm converges rapidly within $100$ training epochs, while the GAI-based algorithm begins to converge after $1000$ training epochs. 
Unfortunately, PPO-based DRL algorithms have low sampling efficiency and tend to be trapped in local optimums, preventing further enhancements in sum SE \cite{du2023beyond}.
On the other hand, our proposed GAI-based algorithm smoothly and continuously progresses towards near-optimal values. The reason is that the diffusion model in our proposed algorithm provides superior sampling quality and demonstrates enhanced capabilities in handling long-term dependencies \cite{du2023beyond}.

Subsequently, the trained GAI model is employed to determine the optimal power-splitting factors and power-control coefficients based on the specified environmental parameters. The resulting sum SE values are depicted in Fig.~\ref{fig:GAI}, offering a comparison with different algorithms.
It is evident that the proposed GAI-based algorithm approaches the expert solution.
When considering the case without RS as a benchmark, the sum SE gain from the GAI-based algorithm consistently exceeds twice that of the heuristic algorithm in the given varying environmental conditions.
This is because the diffusion model accurately learns the mapping relationship between the environment and the optimal solution.
Note that once trained, our proposed GAI-based algorithm exhibits strong generalization capabilities across different environments and provides real-time solutions, making it valuable in practical applications.

\section{Conclusion}\label{se:conclusion}

In this paper, we studied the performance of RS-based CF massive MIMO systems over spatially correlated Rician channels. Taking into account the imperfect CSI caused by pilot contamination and noise, we employ the UatF bound to derive a closed-form expression for the sum SE of the considered systems. Furthermore, the achievable sum SE is calculated as a comparison to confirm our derivation indeed provides a tight lower bound.
It is interesting that employing RS preserves the benefits of the LoS component in boosting the desired signal while simultaneously removing the correspondingly enhanced interference.
Moreover, we proposed low-complexity heuristic algorithms for power-splitting in common messages and power-control in private messages, all based on statistical CSI, ensuring rapid and reliable solutions in large-scale networks.
Furthermore, we developed GAI for the RS-based CF massive MIMO system, enabling the trained diffusion model to directly generate near-optimal power-splitting factors and power-control coefficients based on the given environmental parameters. Significantly, GAI-based algorithms demonstrate robust and efficient performance, particularly in dynamic environments, which we believe greatly enhances the practical applicability of RS-based CF massive MIMO systems.

\newcounter{mytempeqncnt1}
\begin{figure*}[t!]
\normalsize
\setcounter{mytempeqncnt}{1}
\setcounter{equation}{58}
\begin{align}\label{eq:gamma_4}
{\Upsilon _4} &= {\mathbf{\bar h}}_{kl}^{\text{H}}{{{\mathbf{\bar h}}}_{il}}{\mathbf{\bar h}}_{kl}^{\text{H}}{{{\mathbf{\bar h}}}_{jl}} + {\mathbf{\bar h}}_{il}^{\text{H}}{{\mathbf{Q}}_{kl}}{{{\mathbf{\bar h}}}_{jl}} \notag\\
&+ \left\{ {\begin{array}{*{20}{c}}
  { {\mathbf{\bar h}}_{kl}^{\text{H}}{{{\mathbf{\bar Q}}}_{ijl}}{{{\mathbf{\bar h}}}_{kl}}  + {\text{tr}}\left( {{{{\mathbf{\bar Q}}}_{ijl}}{{\mathbf{Q}}_{kl}}} \right) + \left\{ {\begin{array}{*{20}{c}}
  {{\text{tr}}\left( {{{{\mathbf{\bar Q}}}_{kil}}} \right){\text{tr}}\left( {{{{\mathbf{\bar Q}}}_{kjl}}} \right) + {\text{tr}}\left( {{{{\mathbf{\bar Q}}}_{kil}}} \right){\mathbf{\bar h}}_{kl}^{\text{H}}{{{\mathbf{\bar h}}}_{jl}} + {\text{tr}}\left( {{{{\mathbf{\bar Q}}}_{kjl}}} \right){\mathbf{\bar h}}_{kl}^{\text{H}}{{{\mathbf{\bar h}}}_{il}},i \in {\mathcal{P}_k},j \in {\mathcal{P}_i}} \\
  {0,i \notin {\mathcal{P}_k},j \in {\mathcal{P}_i}}
\end{array}} \right.} \\
  {\begin{array}{*{20}{c}}
  {{\text{tr}}\left( {{{{\mathbf{\bar Q}}}_{kil}}} \right){\mathbf{\bar h}}_{kl}^{\text{H}}{{{\mathbf{\bar h}}}_{jl}},i \in {\mathcal{P}_k},j \notin {\mathcal{P}_i}} \\
  {{\text{tr}}\left( {{{{\mathbf{\bar Q}}}_{kjl}}} \right){\mathbf{\bar h}}_{kl}^{\text{H}}{{{\mathbf{\bar h}}}_{il}},i \notin {\mathcal{P}_k},j \in {\mathcal{P}_k},j \notin {\mathcal{P}_i}} \\
  {0,i \notin {\mathcal{P}_k},j \notin {\mathcal{P}_k},j \notin {\mathcal{P}_i}}
\end{array}}
\end{array}} \right.  .
\end{align}
\setcounter{equation}{39}
\hrulefill
\vspace{-0.2cm}
\end{figure*}

\begin{appendices}
\section{Proof of Theorem 1}

With the help of \eqref{eq:r_c} and \eqref{eq:r_p}, we can derive the UatF capacity bound \cite{bjornson2017massive} with
\begin{align}
  {\text{SINR}}_k^{\text{c}} &= {p_{\text{d}}}\mathbb{E}\left\{ {{{\left| {{\text{DS}}_k^{\text{c}}} \right|}^2}} \right\}/\left( {{p_{\text{d}}}\mathbb{E}\left\{ {{{\left| {{\text{REC}}_k^{\text{c}}} \right|}^2}} \right\}} \right. \notag \\
  &\!\!\!\!\! \left. { - {p_{\text{d}}}\mathbb{E}\left\{ {{{\left| {{\text{DS}}_k^{\text{c}}} \right|}^2}} \right\} + \frac{{{p_{\text{d}}}}}{K}\sum\limits_{i = 1}^K {\mathbb{E}\left\{ {{{\left| {{\text{REC}}_{ki}^{\text{p}}} \right|}^2}} \right\}}  + {\sigma ^2}} \right) , \\
    {\text{SINR}}_k^{\text{p}} &= \frac{{{p_{\text{d}}}}}{K}\mathbb{E}\left\{ {{{\left| {{\text{DS}}_k^{\text{p}}} \right|}^2}} \right\}/\left( {\frac{{{p_{\text{d}}}}}{K}\sum\limits_{i = 1}^K {\mathbb{E}\left\{ {{{\left| {{\text{REC}}_{ki}^{\text{p}}} \right|}^2}} \right\}} } \right. \notag \\
  &\left. { - \frac{{{p_{\text{d}}}}}{K}\mathbb{E}\left\{ {{{\left| {{\text{DS}}_k^{\text{p}}} \right|}^2}} \right\} + {\sigma ^2}} \right) .
\end{align}
We compute every term of ${{\text{SINR}}_k^{\text{c}}}$ and ${{\text{SINR}}_k^{\text{p}}}$ to obtain the closed-from sum SE expression of \eqref{eq:sumSE}.

\emph{1) Compute ${\mathbb{E}\left\{ {{\mathbf{g}}_{kl}^{\mathrm{H}}{{{\mathbf{\hat g}}}_{il}}} \right\}}$:} Based on the properties of MMSE estimation, we have ${{{\mathbf{\hat g}}}_{kl}}$ and ${{{\mathbf{\tilde g}}}_{kl}}$ independent and we can derive
\begin{align}
\mathbb{E}\left\{ {{\mathbf{g}}_{kl}^{\text{H}}{{{\mathbf{\hat g}}}_{il}}} \right\} = \mathbb{E}\left\{ {{{\left( {{{{\mathbf{\hat g}}}_{kl}} + {{{\mathbf{\tilde g}}}_{kl}}} \right)}^{\text{H}}}{{{\mathbf{\hat g}}}_{il}}} \right\} = \mathbb{E}\left\{ {{\mathbf{\hat g}}_{kl}^{\text{H}}{{{\mathbf{\hat g}}}_{il}}} \right\} .
\end{align}
With the help of \eqref{Eq:ghat}, we have ${{{\mathbf{\hat h}}}_{kl}}$ and ${{{\mathbf{\hat h}}}_{il}}$ correlated for ${i \in {\mathcal{P}_k}}$, and ${{{\mathbf{\hat h}}}_{kl}}$ and ${{{\mathbf{\hat h}}}_{il}}$ are independent for ${i \notin {\mathcal{P}_k}}$. Therefore, we can derive
\begin{align}\label{eq:gkl_ghatil}
\!\mathbb{E}\!\left\{ {{\mathbf{g}}_{kl}^{\text{H}}{{{\mathbf{\hat g}}}_{il}}} \right\} \!=\! \mathbb{E}\!\left\{ {{\mathbf{\hat g}}_{kl}^{\text{H}}{{{\mathbf{\hat g}}}_{il}}} \right\} \!=\! {\mathbf{\bar h}}_{kl}^{\text{H}}{{{\mathbf{\bar h}}}_{il}} \!+\! \left\{ {\begin{array}{*{20}{c}}
  \!\!\!{{\text{tr}}\left( {{{{\mathbf{\bar Q}}}_{kil}}} \right),i \in {\mathcal{P}_k}} \\
  {0,i \notin {\mathcal{P}_k}}
\end{array}} \right. \!\!.
\end{align}

\emph{2) Compute ${\mathbb{E}\left\{ {{{\left| {{\mathbf{g}}_{kl}^{\mathrm{H}}{{{\mathbf{\hat g}}}_{il}}} \right|}^2}} \right\}}$:} Based on the properties of MMSE estimation that ${{{\mathbf{\hat g}}}_{kl}}$ and ${{{\mathbf{\tilde g}}}_{kl}}$ are independent, we obtain
\begin{align}\label{eq:gklgil}
  \mathbb{E}\left\{ {{{\left| {{\mathbf{g}}_{kl}^{\text{H}}{{{\mathbf{\hat g}}}_{il}}} \right|}^2}} \right\} &= \mathbb{E}\left\{ {{{\left| {{{\left( {{{{\mathbf{\hat g}}}_{kl}} + {{{\mathbf{\tilde g}}}_{kl}}} \right)}^{\text{H}}}{{{\mathbf{\hat g}}}_{il}}} \right|}^2}} \right\} \notag \\
   &= \underbrace {\mathbb{E}\left\{ {{{\left| {{\mathbf{\hat g}}_{kl}^{\text{H}}{{{\mathbf{\hat g}}}_{il}}} \right|}^2}} \right\}}_{{\Upsilon _1}} + \underbrace {\mathbb{E}\left\{ {{{\left| {{\mathbf{\tilde g}}_{kl}^{\text{H}}{{{\mathbf{\hat g}}}_{il}}} \right|}^2}} \right\}}_{{\Upsilon _2}} .
\end{align}
With the help of \eqref{Eq:ghat}, \eqref{Eq:ghat_kl}, \eqref{Eq:gerror_kl} and \cite[Lemma 4 and Lemma 5]{8620255}, we derive
\begin{align}\label{eq:gamma1}
{{\Upsilon _1}} &= {\left| {{\mathbf{\bar h}}_{kl}^{\text{H}}{{{\mathbf{\bar h}}}_{il}}} \right|^2} + {\mathbf{\bar h}}_{il}^{\text{H}}{{\mathbf{Q}}_{kl}}{{{\mathbf{\bar h}}}_{il}} + {\mathbf{\bar h}}_{kl}^{\text{H}}{{\mathbf{Q}}_{il}}{{{\mathbf{\bar h}}}_{kl}} + {\text{tr}}\left( {{{\mathbf{Q}}_{il}}{{\mathbf{Q}}_{kl}}} \right) \notag\\
&+ \left\{ {\begin{array}{*{20}{c}}
  {{{\left| {{\text{tr}}\left( {{{{\mathbf{\bar Q}}}_{kil}}} \right)} \right|}^2} + 2 { {\text{tr}}\left( {{{{\mathbf{\bar Q}}}_{kil}}} \right){\mathbf{\bar h}}_{il}^{\text{H}}{{{\mathbf{\bar h}}}_{kl}} } ,i \in {\mathcal{P}_k}} \\
  {0,i \notin {\mathcal{P}_k}}
\end{array}} \right.  .
\end{align}
Furthermore, we can derive
\begin{align}\label{eq:gamma2}
{\Upsilon _2} = {\mathbf{\bar h}}_{il}^{\text{H}}{{\mathbf{C}}_{kl}}{{{\mathbf{\bar h}}}_{il}} + {\text{tr}}\left( {{{\mathbf{Q}}_{il}}{{\mathbf{C}}_{kl}}} \right) .
\end{align}
Substituting \eqref{eq:gamma1} and \eqref{eq:gamma2} into \eqref{eq:gklgil}, we have
\begin{align}\label{eq:gklgil2}
\mathbb{E}\left\{ {{{\left| {{\mathbf{g}}_{kl}^{\text{H}}{{{\mathbf{\hat g}}}_{il}}} \right|}^2}} \right\} &= {\left| {{\mathbf{\bar h}}_{kl}^{\text{H}}{{{\mathbf{\bar h}}}_{il}}} \right|^2} + {\mathbf{\bar h}}_{il}^{\text{H}}{{\mathbf{R}}_{kl}}{{{\mathbf{\bar h}}}_{il}} \notag\\
&+ {\mathbf{\bar h}}_{kl}^{\text{H}}{{\mathbf{Q}}_{il}}{{{\mathbf{\bar h}}}_{kl}} + {\text{tr}}\left( {{{\mathbf{Q}}_{il}}{{\mathbf{R}}_{kl}}} \right) \notag\\
&\!\!\!\!\!\!\!\!\!\!\!\!\!\!\!\!\!\!\!\! + \left\{ {\begin{array}{*{20}{c}}
  {{{\left| {{\text{tr}}\left( {{{{\mathbf{\bar Q}}}_{kil}}} \right)} \right|}^2} + 2 {{\text{tr}}\left( {{{{\mathbf{\bar Q}}}_{kil}}} \right){\mathbf{\bar h}}_{il}^{\text{H}}{{{\mathbf{\bar h}}}_{kl}}} ,i \in {\mathcal{P}_k}} \\
  {0,i \notin {\mathcal{P}_k}}
\end{array}} \right. .
\end{align}

\emph{3) Compute ${\mathbb{E}\left\{ {{{\left| {{\mathrm{DS}}_k^{\mathrm{c}}} \right|}^2}} \right\}}$:} Using the superposition-based precoding ${{\mathbf{v}}_{{\mathrm{c}},l}} = \sum\nolimits_{i = 1}^K {{{{\mathbf{\hat g}}}_{il}}}$ for common messages and with the help of \eqref{eq:gkl_ghatil}, we obtain
\begin{align}
  \mathbb{E}&\left\{ {{{\left| {{\text{DS}}_k^{\text{c}}} \right|}^2}} \right\} = {\left| {\sum\limits_{l = 1}^L {\sqrt {{\rho _l}{\mu _{{\text{c}},l}}} \sum\limits_{i = 1}^K {\mathbb{E}\left\{ {{\mathbf{g}}_{kl}^{\text{H}}{{{\mathbf{\hat g}}}_{il}}} \right\}} } } \right|^2} \notag \\
   &= {\left| {\sum\limits_{l = 1}^L {\sqrt {{\rho _l}{\mu _{{\text{c}},l}}} \left( {\sum\limits_{i = 1}^K {{\mathbf{\bar h}}_{kl}^{\text{H}}{{{\mathbf{\bar h}}}_{il}}}  + \sum\limits_{i \in {\mathcal{P}_k}}^K {{\text{tr}}\left( {{{{\mathbf{\bar Q}}}_{kil}}} \right)} } \right)} } \right|^2} .
\end{align}

\emph{4) Compute ${\mathbb{E}\left\{ {{{\left| {{\mathrm{REC}}_k^{\mathrm{c}}} \right|}^2}} \right\}}$:} With the help of \cite[Eq. (74)]{10032129}, we first expand it into
\begin{align}\label{eq:rec2}
  &\mathbb{E}\left\{ {{{\left| {{\text{REC}}_k^{\text{c}}} \right|}^2}} \right\} = \sum\limits_{l = 1}^L {{\rho _l}{\mu _{{\text{c}},l}}} \mathbb{E}\left\{ {{{\left| {{\mathbf{g}}_{kl}^{\text{H}}{{\mathbf{v}}_{{\text{c}},l}}} \right|}^2}} \right\} \notag\\
  &+ \sum\limits_{l = 1}^L {\sum\limits_{m \ne l}^L {\sqrt {{\rho _l}{\mu _{{\text{c}},l}}{\rho _m}{\mu _{{\text{c}},m}}} } } \mathbb{E}{\left\{ {{\mathbf{g}}_{kl}^{\text{H}}{{\mathbf{v}}_{{\text{c}},l}}} \right\}^*}\mathbb{E}\left\{ {{\mathbf{g}}_{km}^{\text{H}}{{\mathbf{v}}_{{\text{c}},m}}} \right\}.
\end{align}
Then, substituting the superposition-based common precoding ${{\mathbf{v}}_{{\mathrm{c}},l}} = \sum\nolimits_{i = 1}^K {{{{\mathbf{\hat g}}}_{il}}}$ into \eqref{eq:rec2} to have
\begin{align}
\label{eq:gklvcl2} \mathbb{E}\left\{ {{{\left| {{\mathbf{g}}_{kl}^{\text{H}}{{\mathbf{v}}_{{\text{c}},l}}} \right|}^2}} \right\} &\!=\! \mathbb{E}\left\{ {{{\left| {\sum\limits_{i = 1}^K {{\mathbf{g}}_{kl}^{\text{H}}{{{\mathbf{\hat g}}}_{il}}} } \right|}^2}} \right\} \!=\! \sum\limits_{i = 1}^K {\mathbb{E}\left\{ {{{\left| {{\mathbf{g}}_{kl}^{\text{H}}{{{\mathbf{\hat g}}}_{il}}} \right|}^2}} \right\}}  \notag\\
&+ \sum\limits_{i = 1}^K {\sum\limits_{j \ne i}^K {\underbrace {\mathbb{E}\left\{ {{{\left( {{\mathbf{g}}_{kl}^{\text{H}}{{{\mathbf{\hat g}}}_{il}}} \right)}^*}\left( {{\mathbf{g}}_{kl}^{\text{H}}{{{\mathbf{\hat g}}}_{jl}}} \right)} \right\}}_{{\Upsilon _3}}} } , \\
\label{eq:gklvcl} \mathbb{E}\left\{ {{\mathbf{g}}_{kl}^{\text{H}}{{\mathbf{v}}_{{\text{c}},l}}} \right\} &= \sum\limits_{i = 1}^K {\mathbb{E}\left\{ {{\mathbf{g}}_{kl}^{\text{H}}{{{\mathbf{\hat g}}}_{il}}} \right\}} .
\end{align}
Clearly, the next crucial step is to derive the closed-form expression for ${{\Upsilon _3}}$. Using the properties of independent channel estimate and estimation error from MMSE estimation, we can divide ${{\Upsilon _3}}$ into two components as follows:
\begin{align}\label{eq:gamma_3}
  &{{\Upsilon _3}} = \mathbb{E}\left\{ {{{\left( {{{\left( {{{{\mathbf{\hat g}}}_{kl}} + {{{\mathbf{\tilde g}}}_{kl}}} \right)}^{\text{H}}}{{{\mathbf{\hat g}}}_{il}}} \right)}^*}\left( {{{\left( {{{{\mathbf{\hat g}}}_{kl}} + {{{\mathbf{\tilde g}}}_{kl}}} \right)}^{\text{H}}}{{{\mathbf{\hat g}}}_{jl}}} \right)} \right\}\notag \\
   & = \underbrace {\mathbb{E}\left\{ {{{\left( {{\mathbf{\hat g}}_{kl}^{\text{H}}{{{\mathbf{\hat g}}}_{il}}} \right)}^*}\left( {{\mathbf{\hat g}}_{kl}^{\text{H}}{{{\mathbf{\hat g}}}_{jl}}} \right)} \right\}}_{{\Upsilon _4}} + \underbrace {\mathbb{E}\left\{ {{{\left( {{\mathbf{\tilde g}}_{kl}^{\text{H}}{{{\mathbf{\hat g}}}_{il}}} \right)}^*}\left( {{\mathbf{\tilde g}}_{kl}^{\text{H}}{{{\mathbf{\hat g}}}_{jl}}} \right)} \right\}}_{{\Upsilon _5}} .
\end{align}
Importantly, we then calculate ${{\Upsilon _4}}$ in five cases according to whether UE $k$, UE $i$ and UE $j$ use the same pilot.

(a) For ${i \in {\mathcal{P}_k}}$, ${j \in {\mathcal{P}_k}}$, we have ${{{\mathbf{\hat g}}}_{kl}} = {{{\mathbf{\bar h}}}_{kl}} + {\mathbf{Q}}_{kl}^{\frac{1}{2}}{\mathbf{m}}$, ${{{\mathbf{\hat g}}}_{il}} = {{{\mathbf{\bar h}}}_{il}} + {\mathbf{Q}}_{il}^{\frac{1}{2}}{\mathbf{m}}$, ${{{\mathbf{\hat g}}}_{jl}} = {{{\mathbf{\bar h}}}_{jl}} + {\mathbf{Q}}_{jl}^{\frac{1}{2}}{\mathbf{m}}$ where ${\mathbf{m}} \sim \mathcal{C}\mathcal{N}\left( {{\mathbf{0}},{{\mathbf{I}}_N}} \right)$. Then, we can derive
\begin{align}\label{eq:upsilon4}
   \Upsilon _4^{({\text{a}})} &= \mathbb{E}\left\{ {\left( {\underbrace {{\mathbf{\bar h}}_{kl}^{\text{H}}{{{\mathbf{\bar h}}}_{il}}}_{{a_1}} + \underbrace {{\mathbf{\bar h}}_{kl}^{\text{H}}{\mathbf{Q}}_{il}^{\frac{1}{2}}{\mathbf{m}}}_{{b_1}} + \underbrace {{{\mathbf{m}}^{\text{H}}}{{\left( {{\mathbf{Q}}_{kl}^{\text{H}}} \right)}^{\frac{1}{2}}}{{{\mathbf{\bar h}}}_{il}}}_{{c_1}}} \right.} \right. \notag \\
  &{\left. { + \underbrace {{{\mathbf{m}}^{\text{H}}}{{\left( {{\mathbf{Q}}_{kl}^{\text{H}}} \right)}^{\frac{1}{2}}}{\mathbf{Q}}_{il}^{\frac{1}{2}}{\mathbf{m}}}_{{d_1}}} \right)^*}\left( {\underbrace {{\mathbf{\bar h}}_{kl}^{\text{H}}{{{\mathbf{\bar h}}}_{jl}}}_{{a_2}} + \underbrace {{\mathbf{\bar h}}_{kl}^{\text{H}}{\mathbf{Q}}_{jl}^{\frac{1}{2}}{\mathbf{m}}}_{{b_2}}} \right. \notag \\
  &\left. {\left. { + \underbrace {{{\mathbf{m}}^{\text{H}}}{{\left( {{\mathbf{Q}}_{kl}^{\text{H}}} \right)}^{\frac{1}{2}}}{{{\mathbf{\bar h}}}_{jl}}}_{{c_2}} + \underbrace {{{\mathbf{m}}^{\text{H}}}{{\left( {{\mathbf{Q}}_{kl}^{\text{H}}} \right)}^{\frac{1}{2}}}{\mathbf{Q}}_{jl}^{\frac{1}{2}}{\mathbf{m}}}_{{d_2}}} \right)} \right\} .
\end{align}
Next, each term of \eqref{eq:upsilon4} is calculated as
\begin{align}
\mathbb{E}\left\{ {a_1^*{a_2}} \right\} &= {\mathbf{\bar h}}_{kl}^{\text{H}}{{{\mathbf{\bar h}}}_{il}}{\mathbf{\bar h}}_{kl}^{\text{H}}{{{\mathbf{\bar h}}}_{jl}} , \notag \\
  \mathbb{E}\left\{ {b_1^*{b_2}} \right\} &= \mathbb{E}\!\left\{\! {{\mathbf{\bar h}}_{kl}^{\text{H}}{\mathbf{Q}}_{il}^{\frac{1}{2}}{\mathbf{m}}{{\mathbf{m}}^{\text{H}}}{{\left( {{\mathbf{Q}}_{jl}^{\text{H}}} \right)}^{\frac{1}{2}}}{{{\mathbf{\bar h}}}_{kl}}} \!\right\} \!=\! {\mathbf{\bar h}}_{kl}^{\text{H}}{{{\mathbf{\bar Q}}}_{ijl}}{{{\mathbf{\bar h}}}_{kl}} , \notag \\
  \mathbb{E}\left\{ {c_1^*{c_2}} \right\} &= \mathbb{E}\left\{ {{\mathbf{\bar h}}_{il}^{\text{H}}{\mathbf{Q}}_{kl}^{\frac{1}{2}}{\mathbf{m}}{{\mathbf{m}}^{\text{H}}}{{\left( {{\mathbf{Q}}_{kl}^{\text{H}}} \right)}^{\frac{1}{2}}}{{{\mathbf{\bar h}}}_{jl}}} \right\} \!=\! {\mathbf{\bar h}}_{il}^{\text{H}}{{\mathbf{Q}}_{kl}}{{{\mathbf{\bar h}}}_{jl}} , \notag \\
    \mathbb{E}\left\{ {d_1^*{d_2}} \right\} &= \mathbb{E}\left\{ {{{\mathbf{m}}^{\text{H}}}{{\left( {{\mathbf{Q}}_{kl}^{\text{H}}} \right)}^{\frac{1}{2}}}{\mathbf{Q}}_{il}^{\frac{1}{2}}{\mathbf{m}}{{\mathbf{m}}^{\text{H}}}{{\left( {{\mathbf{Q}}_{jl}^{\text{H}}} \right)}^{\frac{1}{2}}}{\mathbf{Q}}_{kl}^{\frac{1}{2}}{\mathbf{m}}} \right\} \notag \\
   &= {\text{tr}}\left( {{{{\mathbf{\bar Q}}}_{kil}}} \right){\text{tr}}\left( {{{{\mathbf{\bar Q}}}_{kjl}}} \right) + {\text{tr}}\left( {{{{\mathbf{\bar Q}}}_{ijl}}{{\mathbf{Q}}_{kl}}} \right) , \notag
\end{align}
\begin{align}
     \mathbb{E}\left\{ {a_1^*{d_2}} \right\} &=\! \mathbb{E}\!\left\{\! {{\mathbf{\bar h}}_{kl}^{\text{H}}{{{\mathbf{\bar h}}}_{il}}{{\mathbf{m}}^{\text{H}}}{{\left(\! {{\mathbf{Q}}_{kl}^{\text{H}}} \!\right)}^{\frac{1}{2}}}{\mathbf{Q}}_{jl}^{\frac{1}{2}}{\mathbf{m}}} \!\right\} \!=\! {\text{tr}}\!\left(\! {{{{\mathbf{\bar Q}}}_{kjl}}} \!\right)\!{\mathbf{\bar h}}_{kl}^{\text{H}}{{{\mathbf{\bar h}}}_{il}} , \notag \\
  \mathbb{E}\left\{ {d_1^*{a_2}} \right\} &=\! \mathbb{E}\!\left\{\! {{{\mathbf{m}}^{\text{H}}}{{\left(\! {{\mathbf{Q}}_{il}^{\text{H}}} \!\right)}^{\frac{1}{2}}}{\mathbf{Q}}_{kl}^{\frac{1}{2}}{\mathbf{m\bar h}}_{kl}^{\text{H}}{{{\mathbf{\bar h}}}_{jl}}} \!\right\} \!=\! {\text{tr}}\!\left(\! {{{{\mathbf{\bar Q}}}_{kil}}} \!\right)\!{\mathbf{\bar h}}_{kl}^{\text{H}}{{{\mathbf{\bar h}}}_{jl}}  . \notag
\end{align}
The remaining terms are zeros due to the circular symmetry properties of ${\mathbf{m}}$. Then, we derive
\begin{align}\label{eq:gamma_a}
\Upsilon _4^{({\text{a}})} &= {\mathbf{\bar h}}_{kl}^{\text{H}}{{{\mathbf{\bar h}}}_{il}}{\mathbf{\bar h}}_{kl}^{\text{H}}{{{\mathbf{\bar h}}}_{jl}} + {\mathbf{\bar h}}_{il}^{\text{H}}{{\mathbf{Q}}_{kl}}{{{\mathbf{\bar h}}}_{jl}} + {\mathbf{\bar h}}_{kl}^{\text{H}}{{{\mathbf{\bar Q}}}_{ijl}}{{{\mathbf{\bar h}}}_{kl}} \notag\\
&+ {\text{tr}}\left( {{{{\mathbf{\bar Q}}}_{ijl}}{{\mathbf{Q}}_{kl}}} \right) + {\text{tr}}\left( {{{{\mathbf{\bar Q}}}_{kil}}} \right){\text{tr}}\left( {{{{\mathbf{\bar Q}}}_{kjl}}} \right) \notag\\
&+ {\text{tr}}\left( {{{{\mathbf{\bar Q}}}_{kil}}} \right){\mathbf{\bar h}}_{kl}^{\text{H}}{{{\mathbf{\bar h}}}_{jl}} + {\text{tr}}\left( {{{{\mathbf{\bar Q}}}_{kjl}}} \right){\mathbf{\bar h}}_{kl}^{\text{H}}{{{\mathbf{\bar h}}}_{il}} .
\end{align}

(b) For ${i \in {\mathcal{P}_k}}$, ${j \notin {\mathcal{P}_k}}$, we have ${{{\mathbf{\hat g}}}_{kl}} = {{{\mathbf{\bar h}}}_{kl}} + {\mathbf{Q}}_{kl}^{\frac{1}{2}}{\mathbf{m}}$, ${{{\mathbf{\hat g}}}_{il}} = {{{\mathbf{\bar h}}}_{il}} + {\mathbf{Q}}_{il}^{\frac{1}{2}}{\mathbf{m}}$, ${{{\mathbf{\hat g}}}_{jl}} = {{{\mathbf{\bar h}}}_{jl}} + {\mathbf{Q}}_{jl}^{\frac{1}{2}}{\mathbf{m}_1}$ where ${\mathbf{m}_1} \sim \mathcal{C}\mathcal{N}\left( {{\mathbf{0}},{{\mathbf{I}}_N}} \right)$ and is independent with $\mathbf{m}$. Following similar steps in (a), we get
\begin{align}\label{eq:gamma_b}
\Upsilon _4^{({\text{b}})} = {\mathbf{\bar h}}_{kl}^{\text{H}}{{{\mathbf{\bar h}}}_{il}}{\mathbf{\bar h}}_{kl}^{\text{H}}{{{\mathbf{\bar h}}}_{jl}} + {\mathbf{\bar h}}_{il}^{\text{H}}{{\mathbf{Q}}_{kl}}{{{\mathbf{\bar h}}}_{jl}} + {\text{tr}}\left( {{{{\mathbf{\bar Q}}}_{kil}}} \right){\mathbf{\bar h}}_{kl}^{\text{H}}{{{\mathbf{\bar h}}}_{jl}} .
\end{align}

(c) For ${i \notin {\mathcal{P}_k}}$, ${j \in {\mathcal{P}_k}}$, we have ${{{\mathbf{\hat g}}}_{kl}} = {{{\mathbf{\bar h}}}_{kl}} + {\mathbf{Q}}_{kl}^{\frac{1}{2}}{\mathbf{m}}$, ${{{\mathbf{\hat g}}}_{il}} = {{{\mathbf{\bar h}}}_{il}} + {\mathbf{Q}}_{il}^{\frac{1}{2}}{\mathbf{m}_1}$, ${{{\mathbf{\hat g}}}_{jl}} = {{{\mathbf{\bar h}}}_{jl}} + {\mathbf{Q}}_{jl}^{\frac{1}{2}}{\mathbf{m}}$, and derive
\begin{align}\label{eq:gamma_c}
\Upsilon _4^{({\text{c}})} = {\mathbf{\bar h}}_{kl}^{\text{H}}{{{\mathbf{\bar h}}}_{il}}{\mathbf{\bar h}}_{kl}^{\text{H}}{{{\mathbf{\bar h}}}_{jl}} + {\mathbf{\bar h}}_{il}^{\text{H}}{{\mathbf{Q}}_{kl}}{{{\mathbf{\bar h}}}_{jl}} + {\text{tr}}\left( {{{{\mathbf{\bar Q}}}_{kjl}}} \right){\mathbf{\bar h}}_{kl}^{\text{H}}{{{\mathbf{\bar h}}}_{il}} .
\end{align}

(d) For ${i \notin {\mathcal{P}_k}}$, ${j \notin {\mathcal{P}_k}}$, ${j \in \mathcal{P}_i}$, we have ${{{\mathbf{\hat g}}}_{kl}} = {{{\mathbf{\bar h}}}_{kl}} + {\mathbf{Q}}_{kl}^{\frac{1}{2}}{\mathbf{m}}$, ${{{\mathbf{\hat g}}}_{il}} = {{{\mathbf{\bar h}}}_{il}} + {\mathbf{Q}}_{il}^{\frac{1}{2}}{\mathbf{m}_1}$, ${{{\mathbf{\hat g}}}_{jl}} = {{{\mathbf{\bar h}}}_{jl}} + {\mathbf{Q}}_{jl}^{\frac{1}{2}}{\mathbf{m}_1}$, and derive
\begin{align}\label{eq:gamma_d}
\Upsilon _4^{({\text{d}})} &= {\mathbf{\bar h}}_{kl}^{\text{H}}{{{\mathbf{\bar h}}}_{il}}{\mathbf{\bar h}}_{kl}^{\text{H}}{{{\mathbf{\bar h}}}_{jl}} + {\mathbf{\bar h}}_{il}^{\text{H}}{{\mathbf{Q}}_{kl}}{{{\mathbf{\bar h}}}_{jl}} \notag\\
&+ {\mathbf{\bar h}}_{kl}^{\text{H}}{{{\mathbf{\bar Q}}}_{ijl}}{{{\mathbf{\bar h}}}_{kl}} + {\text{tr}}\left( {{{{\mathbf{\bar Q}}}_{ijl}}{{\mathbf{Q}}_{kl}}} \right) .
\end{align}

(e) For ${i \notin {\mathcal{P}_k}}$, ${j \notin {\mathcal{P}_k}}$, ${j \notin \mathcal{P}_i}$, we have ${{{\mathbf{\hat g}}}_{kl}} = {{{\mathbf{\bar h}}}_{kl}} + {\mathbf{Q}}_{kl}^{\frac{1}{2}}{\mathbf{m}}$, ${{{\mathbf{\hat g}}}_{il}} = {{{\mathbf{\bar h}}}_{il}} + {\mathbf{Q}}_{il}^{\frac{1}{2}}{\mathbf{m}_1}$, ${{{\mathbf{\hat g}}}_{jl}} = {{{\mathbf{\bar h}}}_{jl}} + {\mathbf{Q}}_{jl}^{\frac{1}{2}}{\mathbf{m}_2}$, where ${{\mathbf{m}}_2} \sim \mathcal{C}\mathcal{N}\left( {{\mathbf{0}},{{\mathbf{I}}_N}} \right)$ and is independent with ${{\mathbf{m}}}$ and ${{\mathbf{m}}_1}$. We derive
\begin{align}\label{eq:gamma_e}
\Upsilon _4^{({\text{e}})} = {\mathbf{\bar h}}_{kl}^{\text{H}}{{{\mathbf{\bar h}}}_{il}}{\mathbf{\bar h}}_{kl}^{\text{H}}{{{\mathbf{\bar h}}}_{jl}} + {\mathbf{\bar h}}_{il}^{\text{H}}{{\mathbf{Q}}_{kl}}{{{\mathbf{\bar h}}}_{jl}} .
\end{align}
Based on results of \eqref{eq:gamma_a}, \eqref{eq:gamma_b}, \eqref{eq:gamma_c}, \eqref{eq:gamma_d}, \eqref{eq:gamma_e}, we derive $\Upsilon _4$ in \eqref{eq:gamma_4} at the top of this page.
It is worth noting that our derived \eqref{eq:gamma_4} is a more general formula with Lemma 4, and Lemma 5 in \cite{8620255} are special cases of it. It is also the core process of proving Theorem~\ref{thm:1}.
Furthermore, we obtain ${\Upsilon _5}$ as
\setcounter{equation}{59}
\begin{align}\label{eq:gamma_5}
{\Upsilon _5} = {\mathbf{\bar h}}_{il}^{\text{H}}{{\mathbf{C}}_{kl}}{{{\mathbf{\bar h}}}_{jl}} + \left\{ {\begin{array}{*{20}{c}}
  {{\text{tr}}\left( {{{{\mathbf{\bar Q}}}_{ijl}}{{\mathbf{C}}_{kl}}} \right),j \in {\mathcal{P}_i}} \\
  {0,j \notin {\mathcal{P}_i}}
\end{array}} \right. .
\end{align}
Then, substituting \eqref{eq:gklgil2}, \eqref{eq:gamma_3}, \eqref{eq:gamma_4} and \eqref{eq:gamma_5} into \eqref{eq:gklvcl2}, we get
\begin{align}\label{eq:gklvcl22}
  \mathbb{E}&\left\{ {{{\left| {{\mathbf{g}}_{kl}^{\text{H}}{{\mathbf{v}}_{{\text{c}},l}}} \right|}^2}} \right\} = \sum\limits_{i = 1}^K {\sum\limits_{j \in {\mathcal{P}_i}}^K {\left( {{\text{tr}}\left( {{{{\mathbf{\bar Q}}}_{ijl}}{{\mathbf{R}}_{kl}}} \right) + {\mathbf{\bar h}}_{kl}^{\text{H}}{{{\mathbf{\bar Q}}}_{ijl}}{{{\mathbf{\bar h}}}_{kl}}} \right)} }  \notag \\
   &+\! \sum\limits_{i = 1}^K \!{\sum\limits_{j = 1}^K {{\mathbf{\bar h}}_{il}^{\text{H}}{{\mathbf{R}}_{kl}}{{{\mathbf{\bar h}}}_{jl}}} }  \!+\! {\left| {\sum\limits_{i = 1}^K {{\mathbf{\bar h}}_{kl}^{\text{H}}{{{\mathbf{\bar h}}}_{il}}}  \!+\! \sum\limits_{i \in {\mathcal{P}_k}}^K {{\text{tr}}\left( {{{{\mathbf{\bar Q}}}_{kil}}} \right)} } \right|^2} .
\end{align}
Moreover, with the help of \eqref{eq:gkl_ghatil}, \eqref{eq:gklvcl}, \eqref{eq:gklvcl22} and \eqref{eq:rec2}, we derive
\begin{align}
  \mathbb{E}&\left\{ {{{\left| {{\text{REC}}_k^{\text{c}}} \right|}^2}} \right\} = \sum\limits_{l = 1}^L {{\rho _l}} {\mu _{{\text{c}},l}}\left( {\sum\limits_{i = 1}^K {\sum\limits_{j \in {\mathcal{P}_k}}^K {\left( {{\text{tr}}\left( {{{{\mathbf{\bar Q}}}_{ijl}}{{\mathbf{R}}_{kl}}} \right)} \right.} } } \right. \notag \\
  &\left. {\left. { + {\mathbf{\bar h}}_{kl}^{\text{H}}{{{\mathbf{\bar Q}}}_{ijl}}{{{\mathbf{\bar h}}}_{kl}}} \right) + \sum\limits_{i = 1}^K {\sum\limits_{j = 1}^K {{\mathbf{\bar h}}_{il}^{\text{H}}{{\mathbf{R}}_{kl}}{{{\mathbf{\bar h}}}_{jl}}} } } \right) \notag \\
   &+ {\left| {\sum\limits_{l = 1}^L {\sqrt {{\rho _l}{\mu _{{\text{c}},l}}} } \left( {\sum\limits_{i = 1}^K {{\mathbf{\bar h}}_{kl}^{\text{H}}{{{\mathbf{\bar h}}}_{il}}}  + \sum\limits_{i \in {\mathcal{P}_k}}^K {{\text{tr}}\left( {{{{\mathbf{\bar Q}}}_{kil}}} \right)} } \right)} \right|^2} .
\end{align}

\emph{5) Compute ${\mathbb{E}\left\{ {{{\left| {{\mathrm{DS}}_k^{\mathrm{p}}} \right|}^2}} \right\}}$:} Using the MR precoding ${{\mathbf{v}}_{kl}} = {{{\mathbf{\hat g}}}_{kl}}$ for private messages and with the help of \eqref{eq:gkl_ghatil} to derive
\begin{align}
  &\mathbb{E}\left\{ {{{\left| {{\text{DS}}_k^{\text{p}}} \right|}^2}} \right\} = \mathbb{E}\left\{ {{{\left| {\sum\limits_{l = 1}^L {\sqrt {\left( {1 - {\rho _l}} \right){\eta _{kl}}{\mu _{kl}}} } \mathbb{E}\left\{ {{\mathbf{g}}_{kl}^{\text{H}}{{{\mathbf{\hat g}}}_{kl}}} \right\}} \right|}^2}} \right\} \notag \\
   &=  {{{\left| {\sum\limits_{l = 1}^L {\sqrt {\left( {1 - {\rho _l}} \right){\eta _{kl}}{\mu _{kl}}} } \left( {{\mathbf{\bar h}}_{kl}^{\text{H}}{{{\mathbf{\bar h}}}_{kl}} + {\text{tr}}\left( {{{\mathbf{Q}}_{kl}}} \right)} \right)} \right|}^2}}  .
\end{align}

\emph{6) Compute ${\mathbb{E}\left\{ {{{\left| {{\mathrm{REC}}_{ki}^{\mathrm{p}}} \right|}^2}} \right\}}$:} Using the MR private precoding ${{\mathbf{v}}_{il}} = {{{\mathbf{\hat g}}}_{il}}$, we expand ${\mathbb{E}\left\{ {{{\left| {{\mathrm{REC}}_{ki}^{\mathrm{p}}} \right|}^2}} \right\}}$ into
\begin{align}\label{eq:RECpk}
  &\mathbb{E}\left\{ {{{\left| {{\text{REC}}_{ki}^{\text{p}}} \right|}^2}} \right\} = \mathbb{E}\left\{\! {{{\left| {\sum\limits_{l = 1}^L {\sqrt {\left( {1 - {\rho _l}} \right){\eta _{il}}{\mu _{il}}} } {\mathbf{g}}_{kl}^{\text{H}}{{{\mathbf{\hat g}}}_{il}}} \right|}^2}} \!\right\} \notag \\
   &=\! \sum\limits_{l = 1}^L \!{\left( {1 \!-\! {\rho _l}} \right){\eta _{il}}{\mu _{il}}} \mathbb{E}\!\left\{\! {{{\left| {{\mathbf{g}}_{kl}^{\text{H}}{{{\mathbf{\hat g}}}_{il}}} \right|}^2}} \!\right\} \!+\! \sum\limits_{l = 1}^L \!{\sum\limits_{m \ne l}^L \!{\sqrt {\left( {1 \!-\! {\rho _l}} \right){\eta _{il}}{\mu _{il}}} } }  \notag \\
   &\times \sqrt {\left( {1 - {\rho _m}} \right){\eta _{im}}{\mu _{im}}} \mathbb{E}{\left\{ {{\mathbf{g}}_{kl}^{\text{H}}{{{\mathbf{\hat g}}}_{il}}} \right\}^*}\mathbb{E}\left\{ {{\mathbf{g}}_{km}^{\text{H}}{{{\mathbf{\hat g}}}_{im}}} \right\}  .
\end{align}
Finally, plugging \eqref{eq:gkl_ghatil} and \eqref{eq:gklgil2} into \eqref{eq:RECpk}, we derive
\begin{align}
  &\mathbb{E}\left\{ {{{\left| {{\text{REC}}_{ki}^{\text{p}}} \right|}^2}} \right\} = \sum\limits_{l = 1}^L {\left( {1 - {\rho _l}} \right){\eta _{il}}{\mu _{il}}}  \notag \\
   &\times \left( {{\text{tr}}\left( {{{\mathbf{Q}}_{il}}{{\mathbf{R}}_{kl}}} \right) + {\mathbf{\bar h}}_{kl}^{\text{H}}{{\mathbf{Q}}_{il}}{{{\mathbf{\bar h}}}_{kl}} + {\mathbf{\bar h}}_{il}^{\text{H}}{{\mathbf{R}}_{kl}}{{{\mathbf{\bar h}}}_{il}}} \right) \notag \\
   &+\! \left\{ {\begin{array}{*{20}{c}}
  \!\!\!{{{\left| {\sum\limits_{l = 1}^L \! {\sqrt {\left( {1 \!-\! {\rho _l}} \right){\eta _{il}}{\mu _{il}}} \left( {{\mathbf{\bar h}}_{kl}^{\text{H}}{{{\mathbf{\bar h}}}_{il}} \!+\! {\text{tr}}\left( {{{{\mathbf{\bar Q}}}_{kil}}} \right)} \right)} } \right|}^2},i \in {\mathcal{P}_k}} \\
  {{{\left| {\sum\limits_{l = 1}^L {\sqrt {\left( {1 - {\rho _l}} \right){\eta _{il}}{\mu _{il}}} \left( {{\mathbf{\bar h}}_{kl}^{\text{H}}{{{\mathbf{\bar h}}}_{il}}} \right)} } \right|}^2},i \notin {\mathcal{P}_k}}
\end{array}} \right. \!\!. \notag
\end{align}
and this completes the proof.
\end{appendices}

\vspace{0.3cm}
\bibliographystyle{IEEEtran}
\bibliography{IEEEabrv,Ref}

\end{document}